\documentclass[12pt,a4wide]{article}

\usepackage{graphicx}
\usepackage{subcaption} 
\usepackage{caption} 
\usepackage[version=3]{mhchem}
\usepackage{color}
\usepackage{hyperref}

\newtheorem{remark}{Remark}

\newif\ifub 
\ubtrue

\begin{document}

\title{Detection of persistent signals and its relation to coherent feedforward loops}

\author{Chun Tung Chou \\
School of Computer Science and Engineering, \\ University of New South Wales, \\ Sydney, Australia. \\
E-mail: c.t.chou@unsw.edu.au}

\date{\today}

\maketitle

\begin{abstract} 
Many studies have shown that cells use temporal dynamics of signalling molecules to encode information. One particular class of temporal dynamics is persistent and transient signals, i.e. signals of long and short durations respectively. It has been shown that the coherent type-1 feedforward loop with an AND logic at the output (or C1-FFL for short) can be used to discriminate a persistent input signal from a transient one. This has been done by modelling the C1-FFL, and then use the model to show that persistent and transient input signals give, respectively, a non-zero and zero output. {\color{black} The aim of this paper is to make a connection between the statistical detection of a persistent signals and the C1-FFL.} We begin by first formulating a statistical detection problem of distinguishing persistent signals from transient ones. The solution of the detection problem is to compute the log-likelihood ratio of observing a persistent signal to a transient signal. We show that, if this log-likelihood ratio is positive, which happens when the signal is likely to be persistent, then it can be approximately computed by a C1-FFL. Although the capability of C1-FFL to discriminate persistent signals is known, this paper adds an information processing interpretation on how a C1-FFL works as a detector of persistent signals.  
\end{abstract}

\subsubsection*{Keywords:}
Coherent feedforward loops; detection of persistent signals; detection theory; likelihood ratio; dynamical systems; time-scale separation. 


\section{Introduction} 
By analysing the graph of the transcription networks of the bacterium {\sl Escherichia coli} and the yeast {\sl Saccharomyces cerevisiae}, the authors in \cite{Milo:2002cg,ShenOrr:2002jo,Alon:2007uu} discovered that there were sub-graphs that appear much more frequently in these transcription networks than in randomly generated networks. These frequently occurring sub-graphs are called network motifs. A particular example of network motif is the coherent type-1 feedforward loop with an AND logic at the output, or C1-FFL for short. C1-FFL is the most abundant type of coherent feedforward loops in the transcription networks of {\sl E. coli} and {\sl S. cerevisiae} \cite{Mangan:2003ja}. An example of C1-FFL in {\sl E. coli} is the L-arabinose utilisation system which activates the transcription of the araBAD {\color{black}operon} when glucose is absent and L-arabinose is present \cite{Mangan:2003ia}. By modelling the C1-FFL with ordinary differential equations (ODE), the authors in \cite{ShenOrr:2002jo,Mangan:2003ja} show that the C1-FFL can act as persistence detectors to differentiate persistent input signals (i.e. signals of long duration) from transient signals (i.e. signals of short duration). The aim of this paper is to present a new perspective of the persistence detection property of C1-FFL from an information processing point of view. 

In information processing, the problem of distinguishing signals that have some specific features from those that have not, has been studied under the theory of statistical detection ~\cite{Kay_v2}. An approach to detection is to formulate a hypothesis testing problem where the alternative hypothesis (resp. null hypothesis) is that the observed signal does have (does not have) the specific features. The next step is to use the observed signal to compute the likelihood ratio to determine which hypothesis is more likely to hold. Since a C1-FFL can detect persistent signals, a question is whether the C1-FFL can be interpreted as a statistical detector. We show in this paper that the C1-FFL is related to a detection problem whose aim is to distinguish a long rectangular pulse (a prototype persistent signal) from a short rectangular pulse (a prototype transient signal). In particular, we show that, for persistent input signals, the output of the C1-FFL can be interpreted as the log-likelihood ratio of this detection problem. This result therefore provides an information processing interpretation of the computation being carried out by a C1-FFL.

\begin{figure}
    \centering
    \begin{subfigure}[t]{0.4\textwidth}
        \centering
        \includegraphics[scale=0.4]{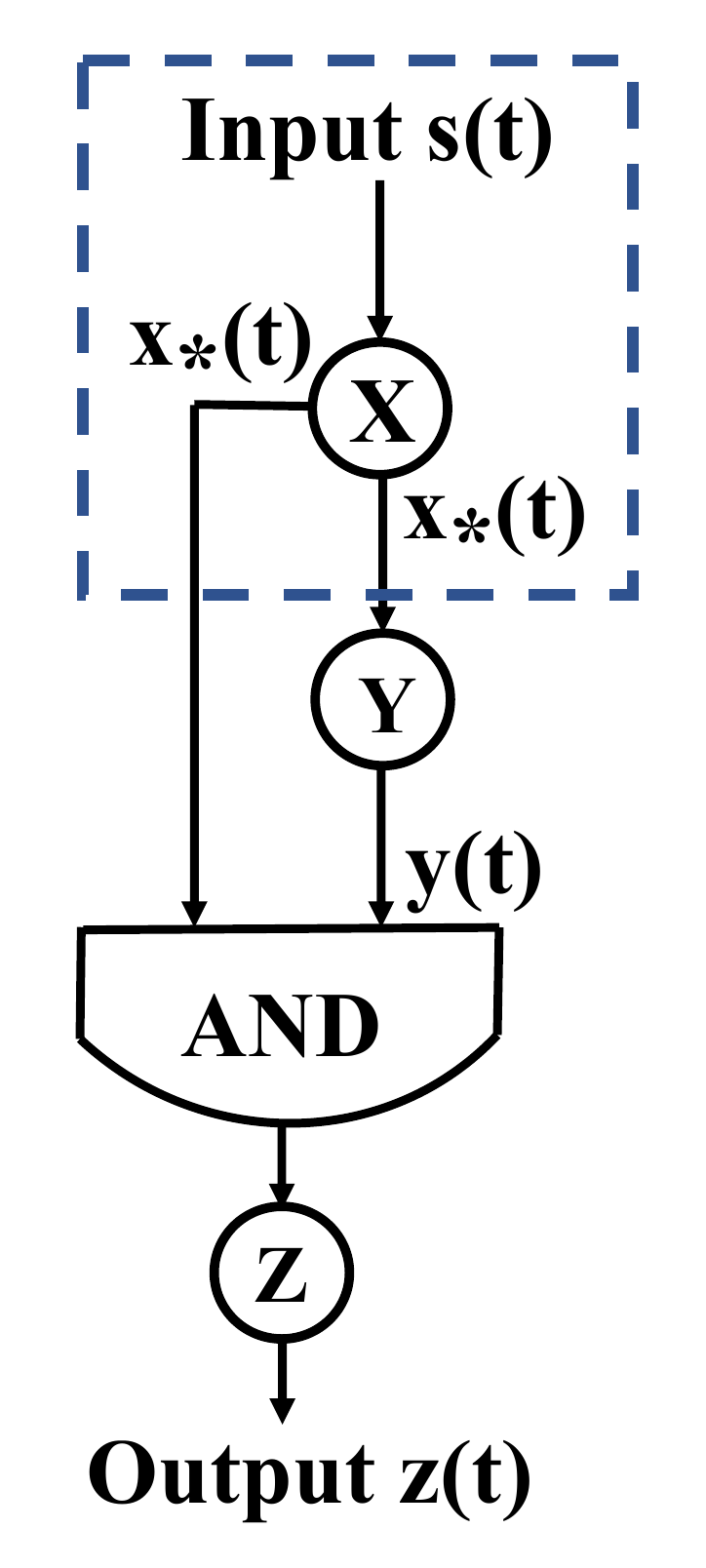}
        \caption{}
        \label{fig:c1ffl}
    \end{subfigure}   
    \begin{subfigure}[t]{0.4\textwidth}
        \centering
        \includegraphics[scale=0.4]{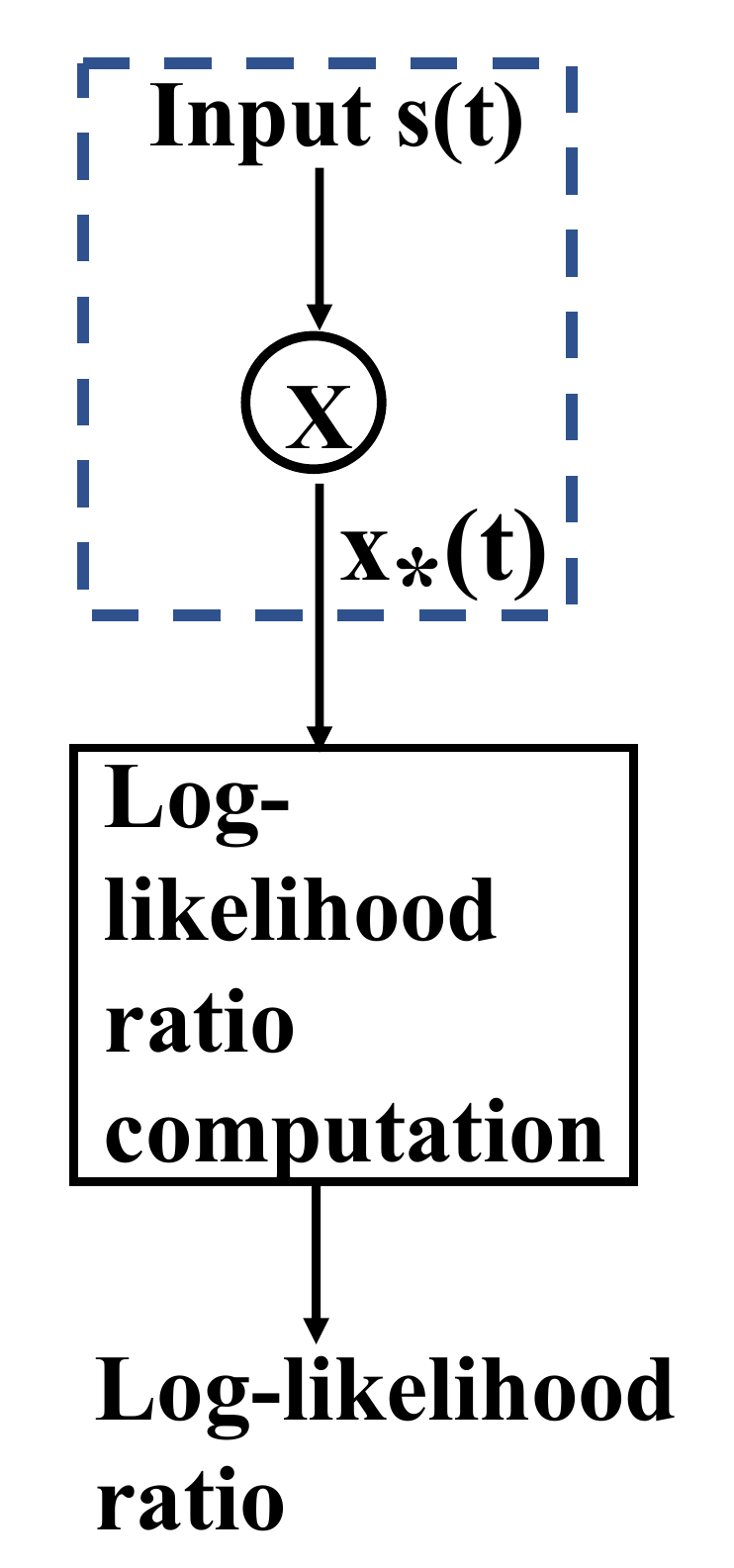}
        \caption{}
        \label{fig:detection theory}
    \end{subfigure}
\caption{(a) The coherent type-1 feedforward loop with AND logic. (b) The detection theory framework.}
\end{figure}

\section{Background}
\label{sec:bg}
\subsection{C1-FFL} 
\label{sec:bg:c1ffl}
The properties of coherent feedforward loops have been studied in \cite{ShenOrr:2002jo,Mangan:2003ja,Alon} using ODE models and in \cite{Mangan:2003ia} experimentally. Here, we will focus on the property of C1-FFL with AND logic to detect persistent signals. We do that by using an idealised model of C1-FFL adapted from the text \cite{Alon}. The model retains the important features of C1-FFL and is useful in understanding the derivation in this paper. \\

Fig.~\ref{fig:c1ffl} depicts the structure of the C1-FFL. One can consider both ${\cee X}$ and ${\cee Y}$ as transcription factors (TFs) which regulate the transcription of ${\cee Z}$. The TF ${\cee X}$ is activated by the input signal $s(t)$ which acts as inducers. We will denote the active form of ${\cee X}$ by ${\cee X_*}$. Following \cite{Mangan:2003ja}, we assume that the activation of ${\cee X}$ (resp. the deactivation of ${\cee X_*}$) is instantaneous when the input signal is present (absent). The active form ${\cee X_*}$ can be used to produce ${\cee Y}$ if its concentration exceeds a threshold $K_{xy}$. We use $[Y]$ to denote the concentration of ${\cee Y}$. We write the reaction-rate equation for ${\cee Y}$ as 
\begin{eqnarray}
\frac{d[Y]}{dt} = \beta_y \; \theta( [{\cee X_*}] > K_{xy}) - \alpha_y [Y]
\label{eq:c1ffl:ideal1}
\end{eqnarray}
where $\beta_y$ and $\alpha_y$ are reaction rate constants, and $\theta(c)$ is 1 if the Boolean condition $c$ within the parentheses is true, and is 0 otherwise. \\

The transcription of ${\cee Z}$ requires the concentration of ${\cee X_*}$ to be greater than $K_{xz}$ {\bf and} the concentration of ${\cee Y}$ to be greater than $K_{yz}$, {\color{black} which corresponds to the AND gate in Fig.~\ref{fig:c1ffl}} . The reaction-rate equation for the output ${\cee Z}$ is: 
\begin{eqnarray}
\frac{d[Z]}{dt} = \beta_z \; \theta( [{\cee X_*}] > K_{xz}) \; \theta( [{\cee Y}] > K_{yz}) - \alpha_z [Z]
\label{eq:c1ffl:ideal2}
\end{eqnarray}
where $\beta_z$ and $\alpha_z$ are reaction rate constants. \\

We now present a numerical example to demonstrate how the C1-FFL can be used to detect persistent signal. We assume the input signal $s(t)$ consists of a short pulse of duration 3 (the transient signal) followed by a long pulse of duration 40 (the persistent signal). We also assume that $s(t)$ has an amplitude of 1 when it is ON. The other parameter values are $\alpha_y = \beta_y = 0.2$, $K_{xy} = 0.6$, $\alpha_z = \beta_z = 1$, $K_{xz} = 0$ and $K_{yz} = 0.5$. \\ 

\begin{figure}
    \centering

\includegraphics[scale=0.5]{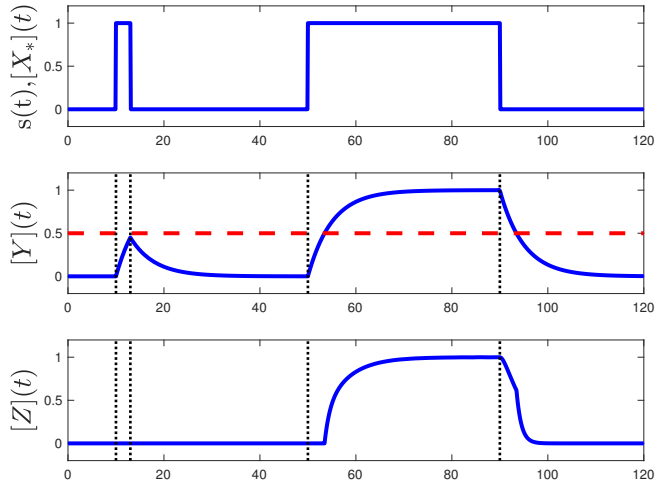}
\captionof{figure}{Illustrating how C1-FFL detects persistent signals.}
\label{fig:bg:persistent}
\end{figure}

Since the activation of ${\cee X}$ or deactivation of ${\cee X_*}$ is instantaneous, we assume $[{\cee X_*}](t) = s(t)$ for simplicity. The time profile of $s(t) = [{\cee X_*}](t)$ is shown in the top plot in Fig.~\ref{fig:bg:persistent}. \\ 

The middle plot of Fig.~\ref{fig:bg:persistent} shows $[Y](t)$. Since $[{\cee X_*}](t) > K_{xy}$ when the input $s(t)$ is ON, the production of ${\cee Y}$ occurs during this period. When the pulse is short, the amount of ${\cee Y}$ being produced is limited and the maximum $[{\cee Y}]$ is below $K_{yz}$, which is indicated by the red horizontal line in the middle plot. {\color{black} Since the production of {\cee Z} requires both $[{\cee X_*}] > K_{xz}$ and  $[{\cee Y}] > K_{yz}$ (i.e. the AND gate) but the latter condition is not satisfied, therefore no ${\cee Z}$ is produced when the pulse is short.} The bottom plot shows $[Z](t)$ is zero when a short pulse is applied. However, when the pulse is long, the concentration of $[{\cee Y}]$ is given enough time to increase beyond the threshold $K_{yz}$ and as a result we see the production of ${\cee Z}$, as shown in the bottom plot. {\color{black} Note that when the pulse of long, the production of ${\cee Z}$ only starts after a delay; this is because the AND condition for the production of {\cee Z} in Eq.~\eqref{eq:c1ffl:ideal2} does not hold initially.} This example shows that, for an ideal C1-FFL, a transient input will produce a zero output and a persistent input will give a non-zero output. 

\subsection{Detection theory} 
Detection theory is a branch of statistical signal processing. Its aim is to use the measured data to decide whether an event of interest has occurred. For example, detection theory is used in radar signal processing to determine whether a target is present or not. In the context of this paper, the events are whether the signal is transient or persistent. A detection problem is often formulated as a hypothesis testing problem, where each hypothesis corresponds to a possible event. Let us consider a detection problem with two hypotheses, denoted by ${\cal H}_0$ and ${\cal H}_1$, which correspond to respectively, the events of transient and persistent signals. Our aim is to decide which hypothesis is more likely to hold. We define the log-likelihood ratio $R$:
\begin{align}
R = \log \left( \frac{{\rm P}[ \mbox{measured data} | {\cal H}_1]}{{\rm P}[\mbox{measured data} | {\cal H}_0]} \right) 
\label{eq:hypo_llr}
\end{align} 
where ${\rm P}[ \mbox{measured data} | {\cal H}_i]$ is the conditional probability that the measured data is generated by the signal specified in hypothesis ${\cal H}_i$. 
Note that we have chosen to use log-likelihood ratio, rather than likelihood ratio, because it will enable us to build a connection with C1-FFL later on.
Intuitively, if the log-likelihood ratio $R$ is positive, then the measured data is more likely to have been generated by a persistent signal or hypothesis ${\cal H}_1$, and vice versa. Therefore, the key idea of detection theory is to use the measured data to compute the log-likelihood ratio and then use it to make a decision. 

\section{Connecting detection theory with C1-FFL} 
We will now present a big picture explanation of how we will connect detection theory with C1-FFL. The signal $x_*(t)$ in Fig.~\ref{fig:c1ffl} is the output signal of Node $X$ in the C1-FFL. We can view the C1-FFL as a 2-stage signal processing engine. In the first stage, the input signal $s(t)$ is processed by Node $X$ to obtain $x_*(t)$ and this is the part within the dashed box in Fig.~\ref{fig:c1ffl}. In the second stage, the signal $x_*(t)$ is processed by the rest of the C1-FFL to produce the output signal $z(t)$. We will now make a connection to detection theory. Our plan is to apply detection theory to the dashed box in Fig.~\ref{fig:c1ffl}.
We consider $x_*(t)$ as the measured data and use them to determine whether the input signal is transient or persistent. Detection theory tells us that we should use $x_*(t)$ to compute the log-likelihood ratio. This means that we can consider the 2-stage signal processing depicted in Fig.~\ref{fig:detection theory} where the input signal $s(t)$ generates $x_*(t)$ and the measured data $x_*(t)$ are used to calculate the log-likelihood ratio. If we can identify the log-likelihood ratio calculation in Fig.~\ref{fig:detection theory} with the processing by the part of C1-FFL outside of the dashed box, then we can identify the signal $z(t)$ with the log-likelihood ratio. 

\section{Detection of persistent signals} 
\label{sec:dp} 

\subsection{Defining the detection problem} 
\label{sec:def:prob}

We first define the problem for detecting a persistent signal using detection theory. Our first step is to specify the signalling pathway in Node $X$, which consists of three chemical species: signalling molecule \cee{S}, molecular type \cee{X} in inactive form and its active form \cee{X_*}. 
The activation and inactivation reactions are: 
\begin{subequations}
\label{cr:all} 
\begin{align}
\cee{
S + X &  ->[k_+] S + X_* \label{cr:on}  \\
X_* &  ->[k_-] X \label{cr:off}}
\end{align}
\end{subequations}
where $k_+$ and $k_-$ are reaction rate constants. Let $x(t)$ and $x_*(t)$ denote, respectively, the {\sl number} of \cee{X} and \cee{X_*} molecules at time $t$. Note that both $x(t)$ and $x_*(t)$ are piecewise constant because they are molecular counts. We assume that $x(t) + x_*(t)$ is a constant for all $t$ and we denote this constant by $M$. 

We assume that the input signal $s(t)$, which is the concentration of the signalling molecules \cee{S} at time $t$, is a deterministic signal. We also assumed that the signal $s(t)$ cannot be observed, so any characteristics of $s(t)$ can only be inferred. 

We model the dynamics of the chemical reactions by using chemical master equation \cite{Gardiner}. This means that $x_*(t)$ is a realisation of a continuous-time Markov chain. This also means that the same input signal $s(t)$ can result in different $x_*(t)$. 

The measured datum at time $t$ is $x_*(t)$. However, in the formulation of the detection problem, we will assume that at time $t$, the data available to the detection problem are $x_*(\tau)$ for all $\tau \in [0,t]$; in other words, the data are continuous in time and are the history of the counts of \cee{X_*} up to time $t$ inclusively. We will use ${\cal X}_*(t)$ to denote the continuous-time history of $x_*(t)$ up to time $t$ inclusively. Note that even though we assume that the entire history ${\cal X}_*(t)$ is available for detection, we will see later on that the calculation of the log-likelihood ratio at time $t$ does not require the storage of the past history. 

The last step in defining the detection problem is to specify the hypotheses ${\cal H}_i$ $(i = 0,1)$. Later on, we will identify ${\cal H}_0$ and ${\cal H}_1$ with, respectively, transient and persistent signals. However, at this stage, we want to solve the detection problem in a general way. We assume that the hypothesis ${\cal H}_0$ (resp. ${\cal H}_1$) is that the input signal $s(t)$ is the signal $c_0(t)$ (resp. $c_1(t)$) where $c_0(t)$ and $c_1(t)$ are two different deterministic signals. Intuitively, the aim of the detection problem is to use the history ${\cal X}_*(t)$ to decide which of the two signals $c_0(t)$ and $c_1(t)$ is more likely to have produced the observed history. 



\subsection{Solution to the detection problem} 

\begin{figure}[t]
    \centering
        \includegraphics[scale=0.35]{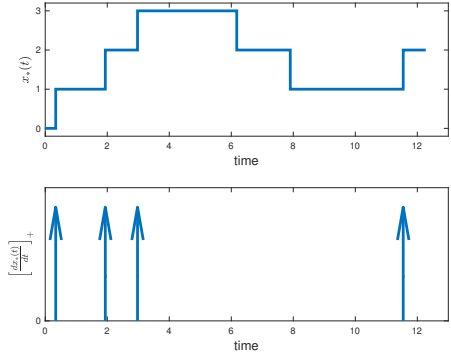}
        \caption{Illustrating $x_*(t)$ and $\left[ \frac{dx_*(t)}{dt} \right]_+$.}
        \label{fig:xstar_deri}
\end{figure} 

Based on the definition of the detection problem, the log-likelihood ratio $L(t)$ at time $t$ is given by:
\begin{align}
L(t) = \log\left( \frac{{\rm P}[{\cal X}_*(t) | {\cal H}_1]}{{\rm P}[{\cal X}_*(t) | {\cal H}_0]} \right)
\label{eq:LLR_t}
\end{align}
where ${\rm P}[{\cal X}_*(t) | {\cal H}_i]$ is the conditional probability of observing the history ${\cal X}_*(t)$ given hypothesis ${\cal H}_i$. We show in Appendix \ref{app:sol:dp} that $L(t)$ obeys the following ODE:
\begin{align}
\frac{dL(t)}{dt} =& \left[ \frac{dx_*(t)}{dt} \right]_+ \log\left(\frac{c_1(t)}{c_0(t)} \right)  - \nonumber  \\ 
 & k_+ (M - x_*(t)) (c_1(t) - c_0(t))        \label{eq:L}
\end{align}
where $[w]_+ = \max(w,0)$. We also assume that the two hypotheses are {\sl a priori} equally likely, so $L(0) = 0$. Since $x_*(t)$ is a piecewise constant function counting the number of \cee{X_*} molecules, its derivative is a sequence of Dirac deltas at the time instants that \cee{X} is activated or \cee{X_*} is deactivated. Note that the Dirac deltas corresponding to the activation of \cee{X} carries a positive sign and the $[  \; ]_+$ operator keeps only these. Figure \ref{fig:xstar_deri} shows an example $x_*(t)$ and its corresponding $\left[ \frac{dx_*(t)}{dt} \right]_+$. {\color{black} We remark that the derivation of \eqref{eq:L} requires that both $c_0(t)$ and $c_1(t)$ are strictly positive for all $t$, otherwise the \eqref{eq:L} is not well defined.}
 
Note that a special case of Eq.~(\ref{eq:L}) with constant $c_i(t)$ and $M = 1$ appeared in \cite{Siggia:2013dd}. An equation of the same form as Eq.~(\ref{eq:L}) is used in \cite{Kobayashi:2011dh} to understand how cells can distinguish between the presence and absence of a stimulus. A more general form of Eq.~(\ref{eq:L}) which includes the diffusion of signalling molecules can be found in \cite{Chou:gc}. 


The importance of Eq.~(\ref{eq:L}) is that, given the measured data $x_*(t)$, we can use it together with $c_i(t)$ to compute the log-likelihood ratio $L(t)$. We will use an example to illustrate how Eq.~(\ref{eq:L}) can be used to distinguish between two signals of different durations. This example will also be used to illustrate what information is useful to distinguish such signals. 

\subsubsection{Example: Using log-likelihood ratio to distinguish between a long and a short pulse} 
\label{sec:dp:ex} 
\begin{figure*}[th]
    \centering
    \begin{subfigure}[t]{0.45\textwidth}
        \centering
        \includegraphics[scale=0.35]{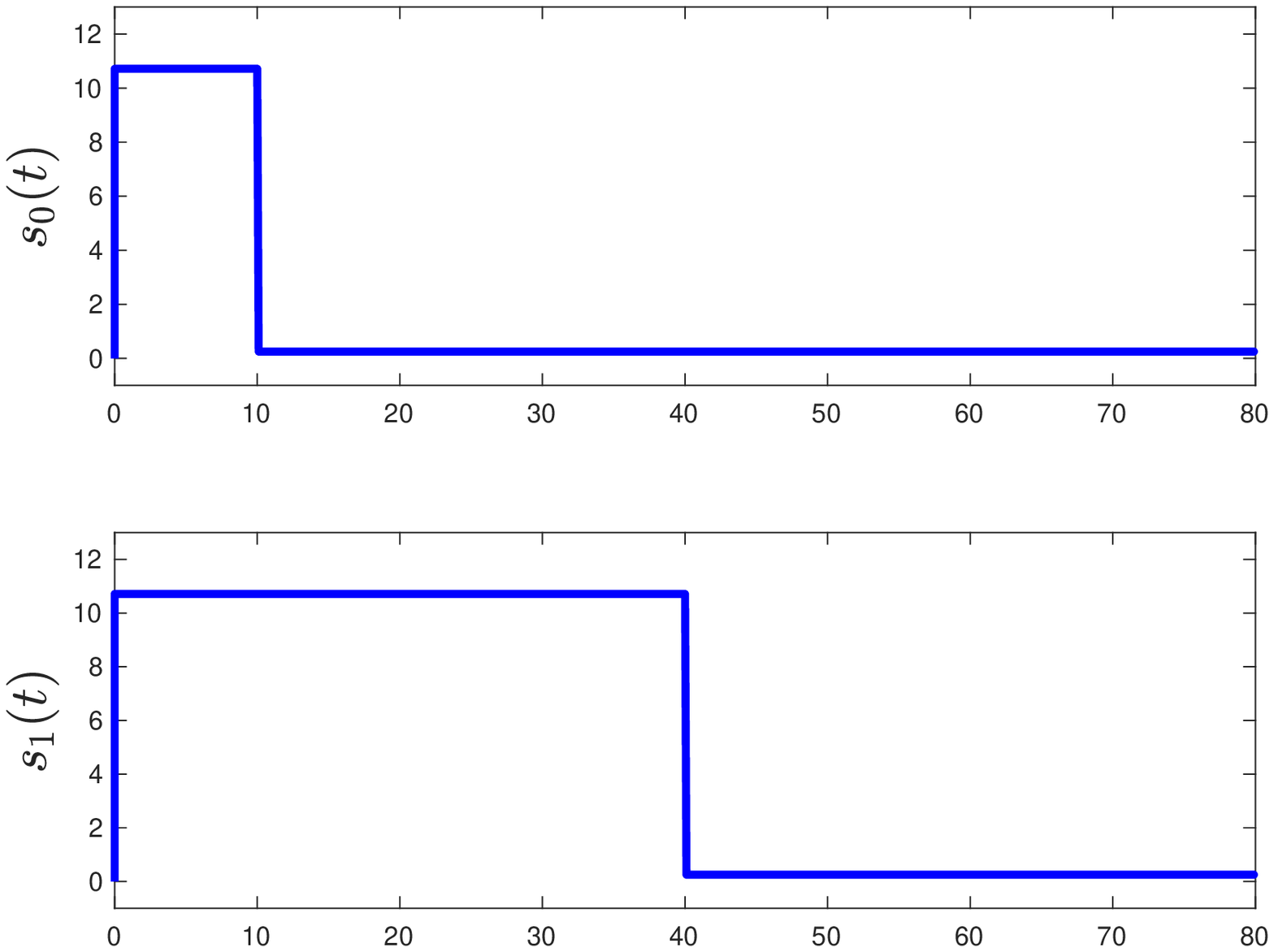}
        \caption{Top graph:  signal $s_0(t)$; Bottom graph: signal $s_1(t)$}
        \label{fig:LLR_signals}
    \end{subfigure}          
    \begin{subfigure}[t]{0.45\textwidth}
        \centering
        \includegraphics[scale=0.35]{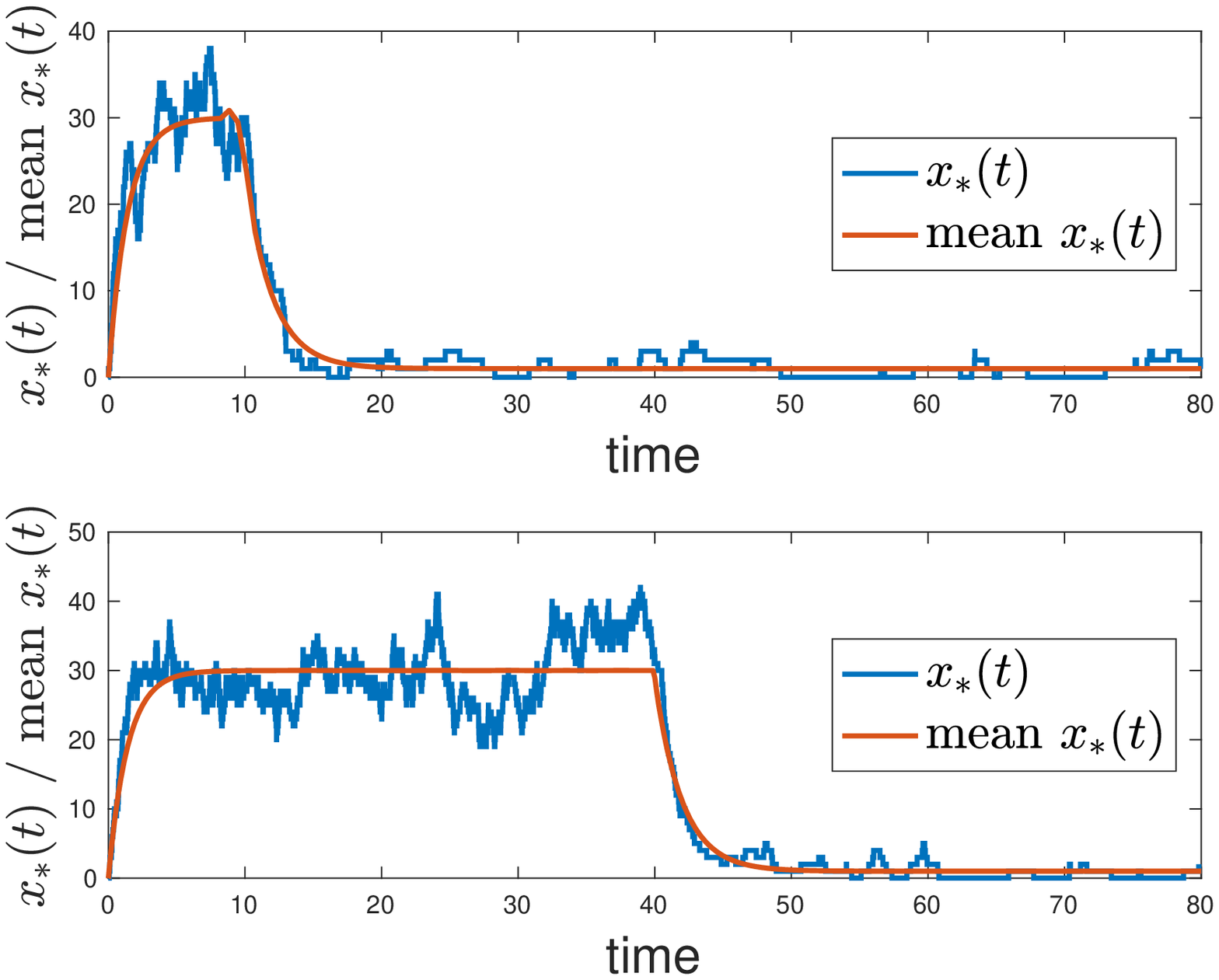}
        \caption{Top graph: $x_*(t)$ for input signal $s_0(t)$; Bottom graph: $x_*(t)$ for input signal $s_1(t)$.}
        \label{fig:xstar}
    \end{subfigure}      
         
    \begin{subfigure}[t]{0.45\textwidth}
        \centering
        \includegraphics[scale=0.35]{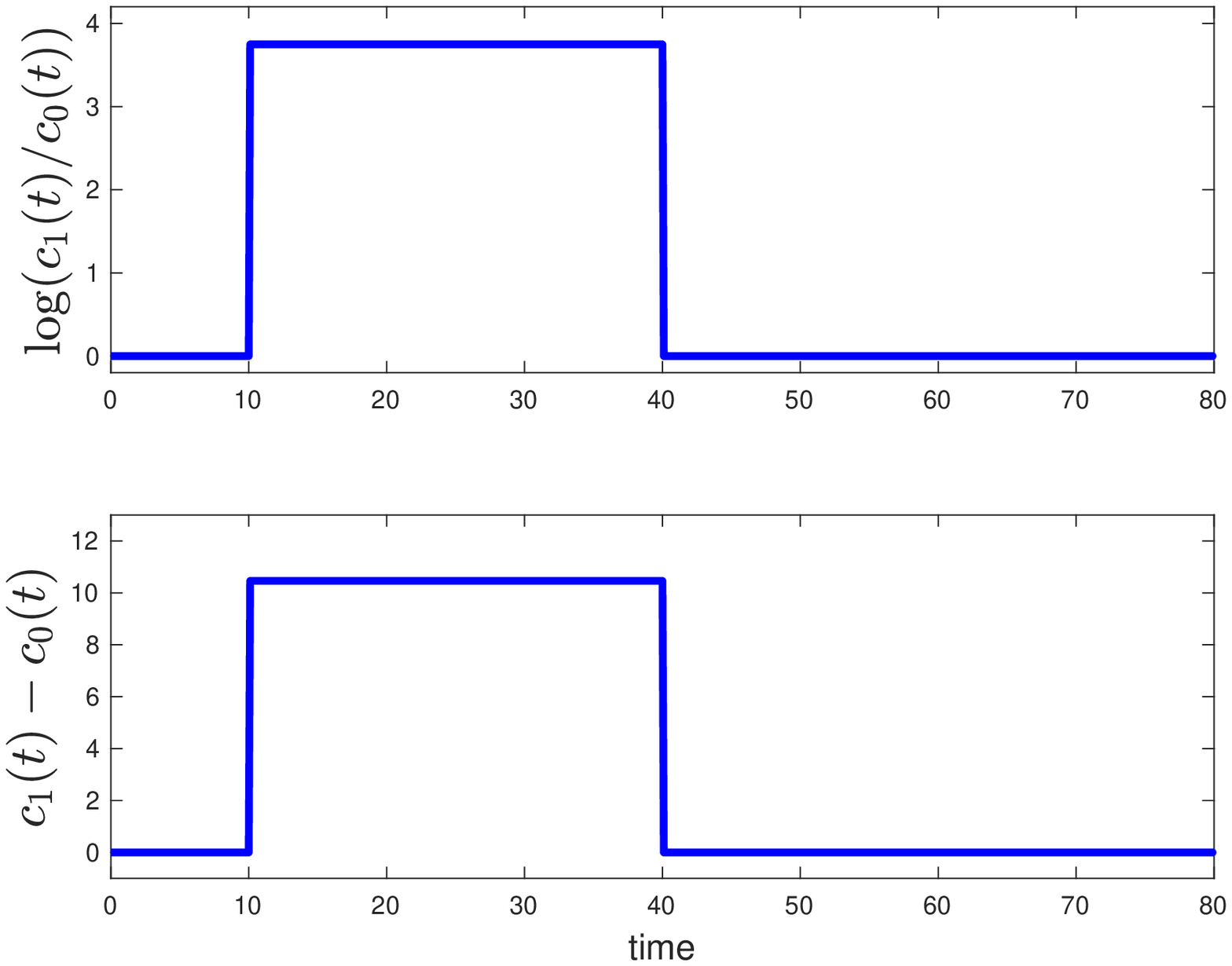}
        \caption{Top graph: $\log(\frac{c_1(t)}{c_0(t)})$; Bottom graph: $c_1(t) - c_0(t)$.}
        \label{fig:weights}
    \end{subfigure}  
     \begin{subfigure}[t]{0.45\textwidth}
        \centering
        \includegraphics[scale=0.35]{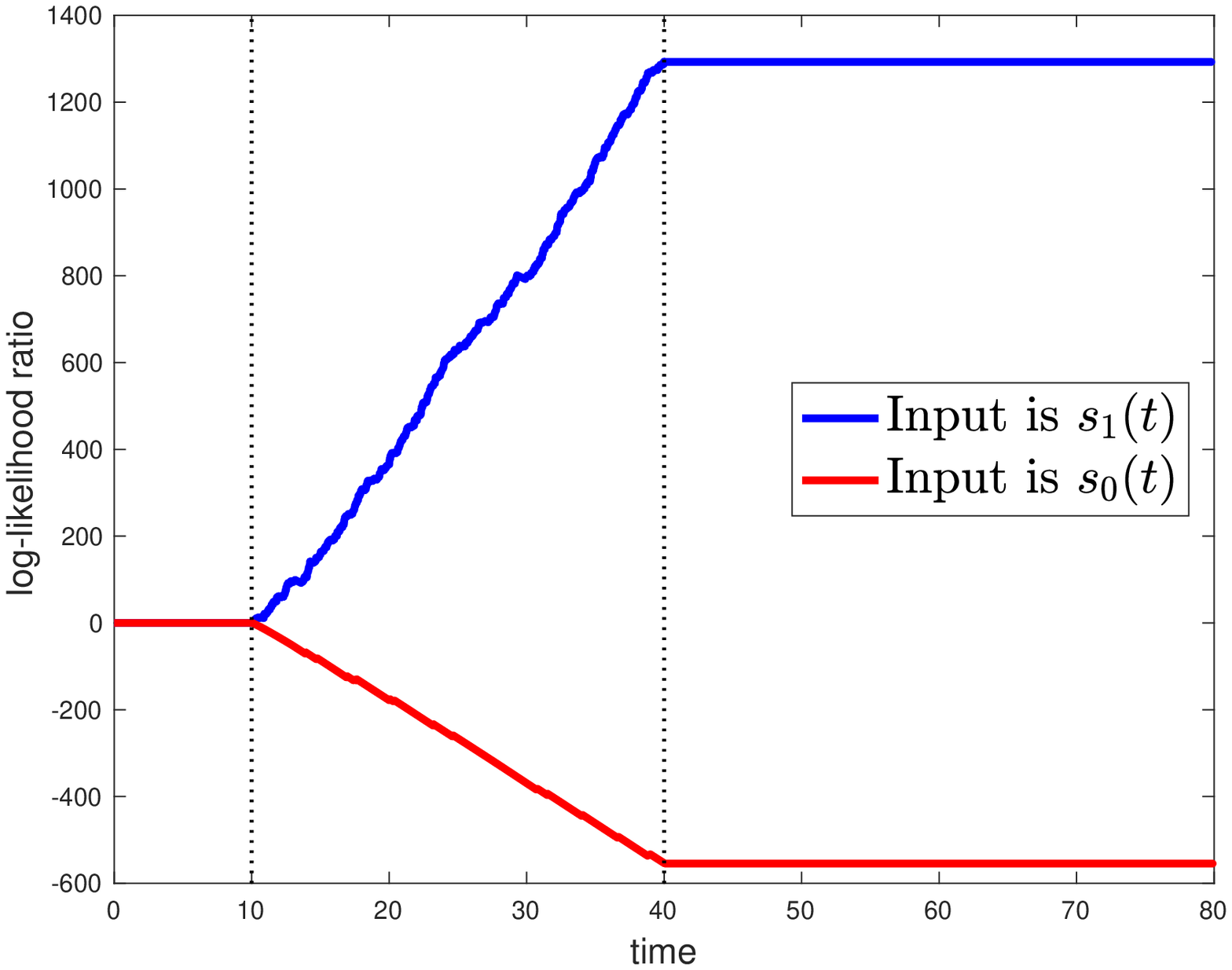}
        \caption{Log-likelihood ratio $L(t)$.}
        \label{fig:LLR_exact}
    \end{subfigure} 
    \caption{Example on distinguishing between a long and a short rectangular pulse. (Best view in colour.)}
\end{figure*}

In this example, we consider using Eq.~(\ref{eq:L}) to distinguish between two possible input signals $s_0(t)$ and $s_1(t)$. Both $s_0(t)$ and $s_1(t)$ are rectangular pulses where $s_1(t)$ has a longer duration than $s_0(t)$. For simplicity, we assume that the reference signals  $c_0(t) = s_0(t)$ and $c_1(t) = s_1(t)$. 

In order to perform the numerical computation, we assume $k_+ = 0.02$, $k_- = 0.5$ and $M = 100$. The time profiles of $s_0(t)$ and $s_1(t)$ are shown in Fig.~\ref{fig:LLR_signals}. The durations of $s_0(t)$ and $s_1(t)$ are, respectively, 10 and 40 time units. The amplitude of the pulses when they are ON is 10.7 and it is 0.25 when they are OFF. 

We use simulation to produce the measured data $x_*(t)$. We first use the input $s_0(t)$ together with the Stochastic Simulation Algorithm \cite{Gillespie:1977ww} to simulate the reactions \eqref{cr:all}. This produces the simulated $x_*(t)$ in the top plot of Fig.~\ref{fig:xstar}. After that, we do the same with $s_1(t)$ as the input and this produces the simulated $x_*(t)$ in the bottom plot of Fig.~\ref{fig:xstar}. It is important to point out that although we have plotted $s_0(t)$, $s_1(t)$ and the two time series of $x_*(t)$ in Figs.~\ref{fig:LLR_signals} and \ref{fig:xstar} using the same time interval, we are doing two separate numerical experiments: one with $s_0(t)$ as the input and the other uses $s_1(t)$ as the input. 

The log-likelihood ratio calculation in Eq.~(\ref{eq:L}) uses the reference signals $c_0(t)$ and $c_1(t)$. We see from Eq.~(\ref{eq:L}) that these two reference signals are used to form two weighting functions $\log\left(\frac{c_1(t)}{c_0(t)} \right)$ and $(c_1(t) - c_0(t))$. By using the assumed time profiles of $c_0(t)$ and $c_1(t)$, we can compute these two weighting functions and we have plotted them in Fig.~\ref{fig:weights}. It can be seen that both weighting functions are non-zero in the time interval $[10,40)$ but zero otherwise. This means that the computation of $L(t)$ is only using the measured data in the time interval $[10,40)$ to determine whether the input signal is $c_0(t)$ or $c_1(t)$. This is because, outside of the time interval $[10,40)$, the two data series $x_*(t)$ generated by $s_0(t)$ and $s_1(t)$ have the same statistical behaviour and therefore there is no information outside of $[10,40)$ to say whether the input is long or short. Hence, a lesson we have learnt from this example is that the informative part of the data is when the long pulse is expected to be ON and the short pulse is expected to be OFF.

We first use the $x_*(t)$ generated by $s_0(t)$, together with the time profiles of $c_0(t)$ and $c_1(t)$, to compute the log-likelihood ratio $L(t)$ by numerically integrating Eq.~(\ref{eq:L}). The resulting $L(t)$ is the red curve in Fig.~\ref{fig:LLR_exact}. Similarly, the blue curve in Fig.~\ref{fig:LLR_exact} shows the $L(t)$ corresponding to the input $s_1(t)$. We can see distinct behaviours in the two $L(t)$'s in the time intervals $[0,10)$, $[10,40)$ and $t \geq 40$. The behaviour in the time intervals $[0,10)$ and $t \geq 40$ is simple to explain because $\frac{dL}{dt} = 0$ in these time intervals.  


We next focus on the time interval $[10,40)$. We first consider $s_1(t)$ as the input. In this time interval, a large $s_1(t)$ means the activation ${\cee X}$ continues to happen, see the bottom plot of Fig.~\ref{fig:xstar}. The activation of ${\cee X}$ contributes to an increase in $L(t)$ due to the first term on the right-hand side (RHS) of Eq.~(\ref{eq:L}). Although the second term of Eq.~(\ref{eq:L}) contributes to a decrease in $L(t)$ via $(M-x_*(t))$, which is the number of inactive X, the contribution is small comparatively. Therefore, we see that the log-likelihood ratio $L(t)$, which is the blue curve in Fig.~\ref{fig:LLR_exact}, becomes more positive. Since a positive log-likelihood ratio means that the input signal is more likely to be similar to the reference signal $c_1(t)$, this is a correct detection. In a similar way, we can explain the behaviour of the red curve in Fig.~\ref{fig:LLR_exact} when $s_0(t)$ is applied. 

A lesson that we can learn from the last paragraph is that, if our aim is to distinguish a persistent signal from a transient one accurately, then we want the persistent signal to produce a large positive $L(t)$. Since the positive contribution of $L(t)$ comes from the first term on the RHS of Eq.~\eqref{eq:L}, we can get a large positive $L(t)$ by making sure that a persistent signal will produce many activations. This occurs when a persistent signal has a duration which is long compared to time scale of the activation and deactivation reactions (\ref{cr:all}) and we will make use of this condition later. 


\section{Connecting log-likelihood calculation to C1-FFL} 

\subsection{Choosing detection problem parameters to match the behaviour of C1-FFL}
The detection problem defined in Section \ref{sec:dp} is general and can be applied to any two chosen reference signals $c_0(t)$ and $c_1(t)$. In order to connect the detection problem in Section \ref{sec:dp} to the fact that C1-FFL is a persistence detector, we will need to make specific choices for $c_0(t)$ and $c_1(t)$. In this paper, we will choose the reference signals $c_0(t)$ and $c_1(t)$ to be rectangular (or ON/OFF) pulses. Furthermore, we assume that when the reference signal is ON, its concentration level is $a_1$; and when it is OFF, its concentration level is at the basal level $a_0$ with $a_1 > a_0 > 0$. The temporal profile of $c_i(t)$ (where $i = 0, 1$) is: 
\begin{eqnarray}
c_i(t) & = & 
\left\{
\begin{array}{ll}
a_1  & \mbox{for } 0 \leq t < d_i \\
a_0 & \mbox{otherwise} 
\end{array}
\right.
\end{eqnarray} 
where $d_i$ is the duration of the pulse $c_i(t)$. In particular, we assume that the duration of $c_1(t)$ is longer than $c_0(t)$, i.e. $d_1 > d_0$. We can therefore identify  $c_0(t)$ and $c_1(t)$ as the reference signals for, respectively, the transient and persistent signals. 

We remark that there may be other choices of reference signals that can connect the detection problem in Section \ref{sec:dp} to the one solved by C1-FFL, we will leave that for future work. 

{\color{black} 
\begin{remark}
We would like to make a remark on the detection problem formulation. 
In this paper we have chosen to formulate the detection problem by assuming that each hypothesis ${\cal H}_0$ and ${\cal H}_1$ consists of one reference signal. Such hypotheses, which consist of only one possibility per hypothesis, are known as simple hypotheses in the statistical hypothesis testing literature \cite{Kay_v2}. We know from \cite{Kay_v2} that if both hypotheses are simple, then the solution of the detection problem is to compute the  likelihood ratio \eqref{eq:hypo_llr}. 
In this paper, we have chosen to use simple hypotheses for ${\cal H}_0$ and ${\cal H}_1$ so as to make the problem trackable. In order to understand that, let us explore an alternative detection problem formulation. 

An alternative formulation would be to assume that ${\cal H}_0$ (resp.~${\cal H}_1$) consists of all rectangular pulses with duration less than (greater than or equal to) a pre-defined threshold $d_0$. In this case, both ${\cal H}_0$ and ${\cal H}_1$ are known as composite hypotheses. To the best of our knowlesge, there are no standard solutions to the hypothesis testing problem with composite hypotheses at the moment. Although the text \cite{Kay_v2} presented two methods to deal with composite hypotheses, neither of them appears to be trackable because the Bayesian approach requires the evaluation of an integral and the generalised likelihood ratio test requires the solution to two optimisation problems. We have therefore not considered them in this paper. 

\end{remark}
}


\subsection{Computing an intermediate approximation} 
\label{sec:conn:ia}
Our ultimate goal is to connect the computation of the log-likelihood ratio $L(t)$ in Eq.~(\ref{eq:L}) to the computation carried out by C1-FFL. We will first derive an intermediate approximation for Eq.~(\ref{eq:L}). In order to motivate why this intermediate approximation is necessary, one first needs to know that the C1-FFL realises computation by using chemical reactions and research from molecular computation in synthetic biology has taught us that some computations are difficult to be carried out by chemical reactions \cite{Oishi:2011ig}. For Eq.~(\ref{eq:L}), the difficulties are: (1) The log-likelihood ratio can take any real value but chemical concentration can only be non-negative; (2) It is difficult to calculate derivatives using chemical reactions. The aim of the intermediate approximation is to remove these difficulties. {{\color{black}In addition, we want the computation to make use of $x_*(t)$ (number of active species ${\cee X_*}$) instead of $M-x_*(t)$ (number of inactive species ${\cee X}$) because signalling pathways typically use the active species to propagate information.}

In order to analytically derive the intermediate approximation, we will need to assume that the input signal $s(t)$ has a certain form. Our derivation assumes that the input $s(t)$ is a rectangular pulse with the following temporal profile: 
\begin{eqnarray}
s(t) & = & 
\left\{
\begin{array}{ll}
a  & \mbox{for } 0 \leq t < d \\
a_0 & \mbox{otherwise} 
\end{array}
\right.
\label{eq:def_pulse_st}
\end{eqnarray} 
where $d$ is the pulse duration, and $a$ is the pulse amplitude when it is ON where $a > a_0$.  Note that the parameters $a$ and $d$ are not fixed; and we will show that the intermediate approximation holds for a range of $a$ and $d$. 

In Appendix \ref{app:Lhat}, we start from Eq.~\eqref{eq:L} and use a time-scale separation argument to derive the intermediate approximation $\hat{L}(t)$. 
{\color{black} The intermediate approximation $\hat{L}(t)$ has the following properties: if the input signal $s(t)$ is persistent, then $\hat{L}(t)$ approximates the log-likelihood ratio $L(t)$; if the input signal $s(t)$ is transient, then $\hat{L}(t)$ is zero. Note that the latter property is consistent with the behaviour of the ideal C1-FFL which gives a zero output for transient signals.} The time evolution of $\hat{L}(t)$ is given by the following ODE: 
\begin{eqnarray}
\frac{d\hat{L}(t)}{dt} &=& \; x_*(t) \times \underbrace{\{ k_- \; \pi(t) \;  [\phi(s(t))]_+ \}}_{= \eta(t)}     \label{eq:Lfinal} \\ 
\mbox{ where } 
\phi(u) &=& \log\left(\frac{a_1}{a_0} \right)  -  \frac{ a_1 - a_0 }{u},    \label{eq:phi} \\
\pi(t) &=& \left\{
\begin{array}{cl}
1 & \mbox{for } d_0 \leq t < d_1 \\
0 & \mbox{otherwise}
\end{array}
\right.  \\
\hat{L}(0) &=& 0 
\end{eqnarray}

The behaviour of the intermediate approximation $\hat{L}(t)$ depends on the duration $d$ of the input signal $s(t)$. Two important properties for $\hat{L}(t)$, which are discussed in further details in Appendix \ref{app:Lhat}, are:
\begin{enumerate}
\item If $d < d_0$, then $\hat{L}(t)$ is zero for all $t$. 
\item If $d \geq d_0$ and if the duration $d - d_0$ is long compared to $\frac{1}{k_+ a + k_-}+\frac{1}{k_-}$, then $\hat{L}(t) \approx L(t)$ for $0 \leq t < \min\{d,d_1\}$ where $L(t)$ is given in Eq.~\eqref{eq:L}. 
\end{enumerate} 

We can consider those input signals $s(t)$ whose duration $d$ is less than $d_0$ as transient signals. The first property says that these signals will give a zero $\hat{L}(t)$. Note that for the ideal C1-FFL considered in Section \ref{sec:bg:c1ffl}, a transient signal gives a zero output. 

Those signals whose duration $d$ is greater than or equal to $d_0$ are considered to be persistent. The second property concerns persistent signals with the property that the  duration $d$ and amplitude $a$ have to be such that $d-d_0$ is long compared to $\frac{1}{k_+ a + k_-}+\frac{1}{k_-}$, which is the mean time between two consecutive activations of an ${\cee X}$ molecule. The physical effect of these signals is to produce a large number of activations and deactivations when the input signal $s(t)$ is ON. We argue in Appendix \ref{app:Lhat} that, if these conditions hold, then it is possible to use $\hat{L}(t)$ in \eqref{eq:Lfinal} to approximate the log-likelihood $L(t)$ in the time interval $0 \leq t < \min\{d,d_1\}$. 


We discussed in Section \ref{sec:dp:ex} that the detection of a persistent signal is best if there are many activations and deactivations when the persistent signal is ON. Fortunately, this is exactly the condition required for the second property to hold. {\color{black} Note that in the analysis of the ideal C1-FFL in \cite{ShenOrr:2002jo,Mangan:2003ja,Alon} and in Section \ref{sec:bg:c1ffl}, both the activation and deactivation reactions \eqref{cr:all} are assumed to be instantaneous, which can be viewed as $k_+$ and $k_-$ being very large. This assumption can be justified from the fact that for C1-FFL, the molecule species ${\cee S}$ and ${\cee X}$ can be considered to be, respectively, an inducer and a transcription factor. It is known that the activation and deactivation dynamics of transcription factors are fast, see \cite[Table 2.1]{Alon}. Hence this assumption is not stringent and we will assume that reactions \eqref{cr:all} are fast for the rest of this paper. } 

We remark that the second property does not cover all the persistent signals. For example, signals with a small amplitude $a$ which do not produce large enough number of activations and inactivations are not covered. These signals are persistent but are hard to detect. 

In the beginning of this section, we mentioned some difficulties in realising the computation of $L(t)$ in Eq.~\eqref{eq:L} using chemical reactions. We note that those difficulties are no longer present in the computation of $\hat{L}(t)$ using \eqref{eq:Lfinal}. In particular, we note that $\hat{L}(t)$ is always non-negative and can be interpreted as log-likelihood ratio when the input is persistent. 

\subsubsection{Numerical illustration} 
\label{sec:lhat:eval} 

\setcounter{subfigure}{0}
\begin{figure}[t]
    \begin{subfigure}[t]{0.45\textwidth}
        \centering
        \includegraphics[scale=0.30]{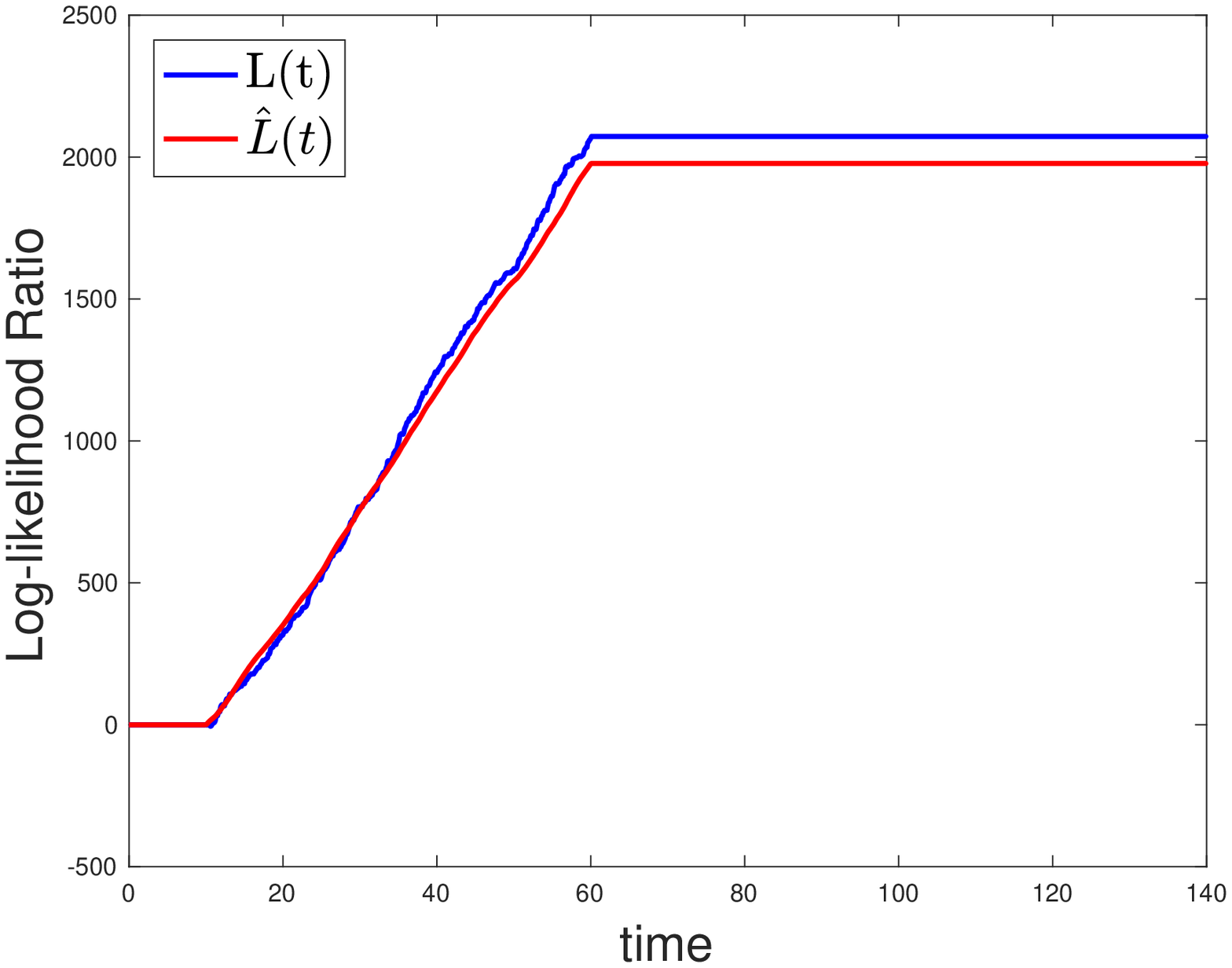}
        \caption{} 
        \label{fig:LLR_approx1_2curves}
    \end{subfigure} 
    \begin{subfigure}[t]{0.45\textwidth}
        \centering
        \includegraphics[scale=0.30]{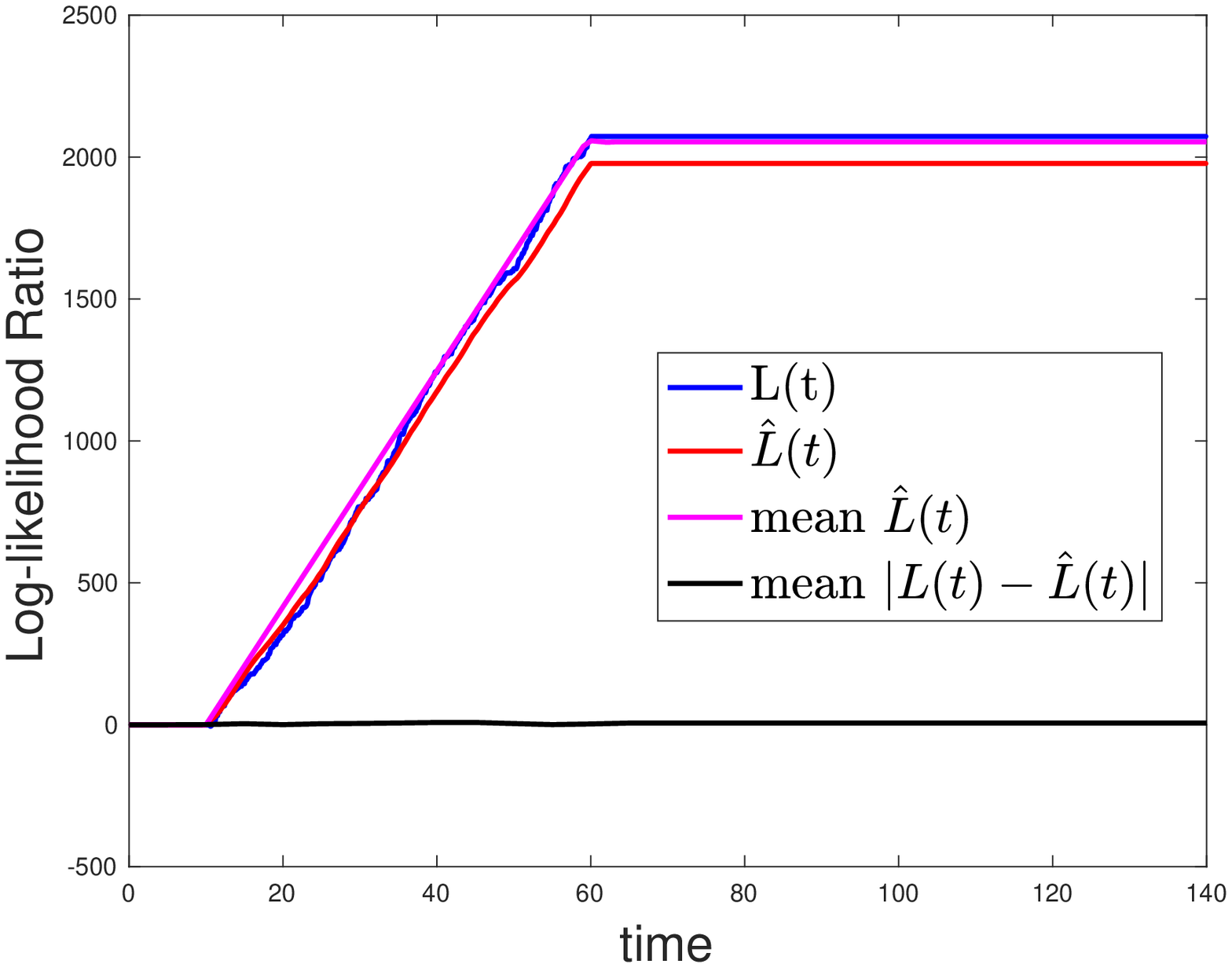}
        \caption{} 
        \label{fig:LLR_approx_1}
    \end{subfigure} 
    
    \begin{subfigure}[t]{0.45\textwidth}
        \centering
        \includegraphics[scale=0.30]{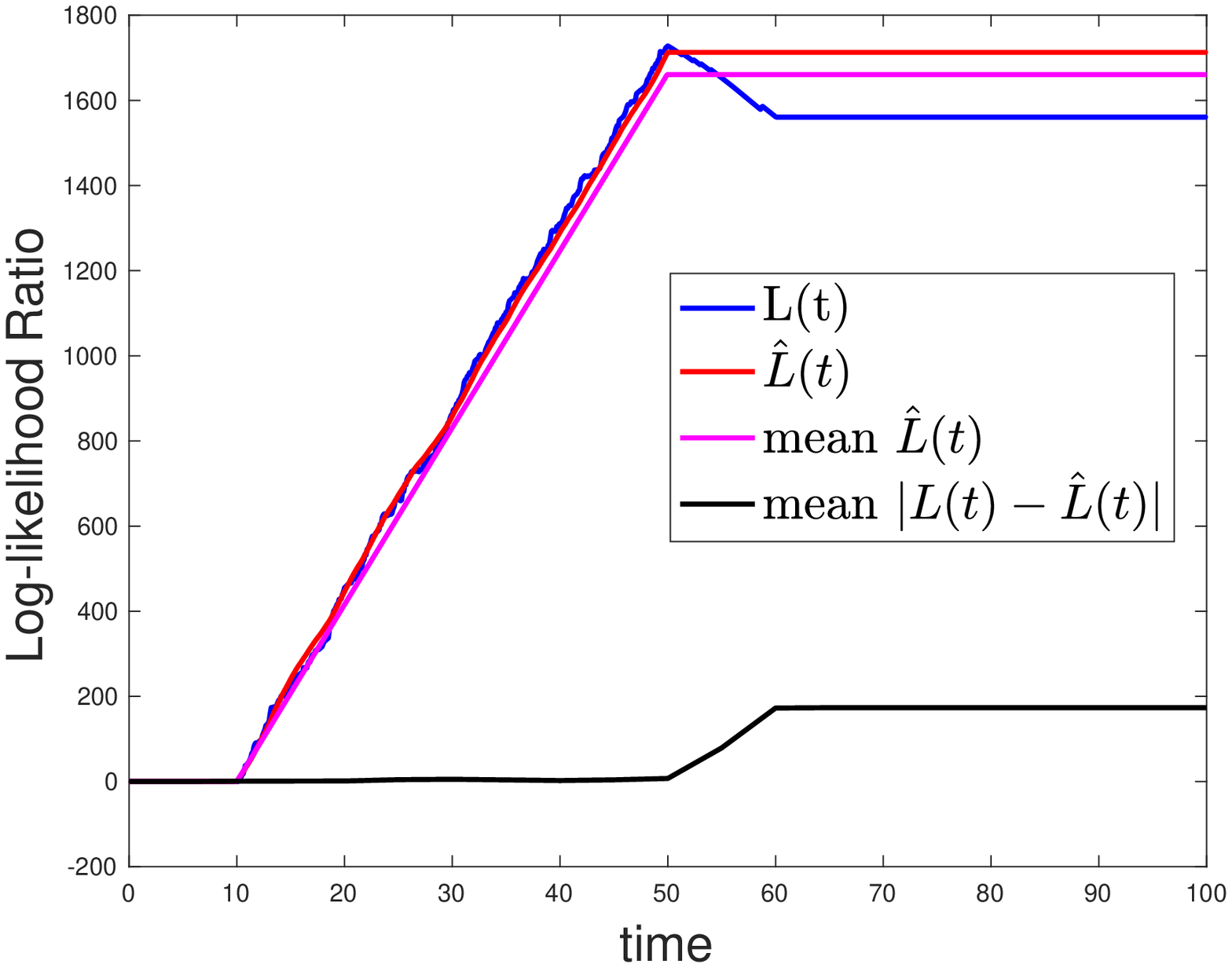}
        \caption{} 
        \label{fig:LLR_approx_2}
    \end{subfigure} 
    \begin{subfigure}[t]{0.45\textwidth}
        \centering
        \includegraphics[scale=0.30]{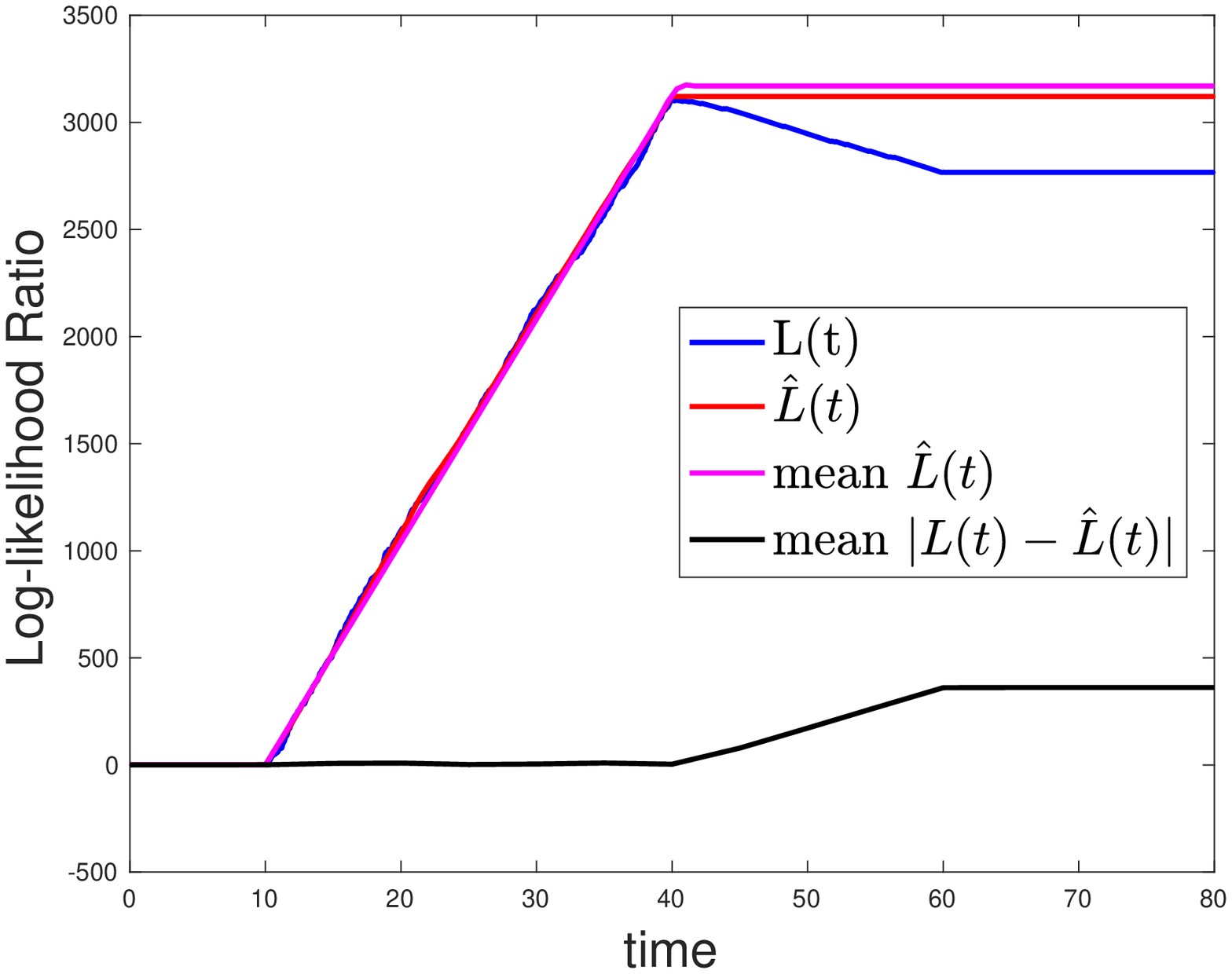}
        \caption{} 
        \label{fig:LLR_approx_3}
    \end{subfigure}     
\caption{Numerical results for intermediate approximation. (Best view in color.)}
\label{fig:LLR_approx}
\end{figure}

We will now use a few numerical examples to illustrate that the intermediate approximation $\hat{L}(t)$ is approximately equal to the log-likelihood ratio $L(t)$ for persistent signals. For all these examples, we choose $k_+=0.02$, $k_-=0.5$, $d_0 = 5$, $d_1 = 60$, $a_0 = 0.25$ and $a_1= 10.7$. 


For the first example, we choose $d = 70$ and $a = a_1$ for the input signal $s(t)$. We use the Stochastic Simulation Algorithm to obtain a realisation of $x_*(t)$. We then use $x_*(t)$ to compute $L(t)$ and $\hat{L}(t)$. The results are shown in Fig.~\ref{fig:LLR_approx1_2curves}. We can see that the approximation is good. We next generate 100 different realisations of $x_*(t)$ and use them to compute $L(t)$ and $\hat{L}(t)$. Fig.~\ref{fig:LLR_approx_1} shows the mean of $|L(t) - \hat{L}(t)|$ over 100 realisations, as well as one realisation of $L(t)$ and $\hat{L}(t)$. It can be seem that the approximation error is small. In Fig.~\ref{fig:LLR_approx_1}, we have also plotted the mean of $\hat{L}(t)$ obtained by solving the following system of ODEs: 
\begin{eqnarray}
\frac{d\bar{x}_*(t)}{dt} &=& k_+ s(t) (M - \bar{x}_*(t)) - k_- \bar{x}_*(t) \label{eq:xstar_bar} \\
\frac{d\bar{L}(t)}{dt} &=& \; \bar{x}_*(t) \times  k_- \; \pi(t) \;  [\phi(s(t))]_+ 
\end{eqnarray}
where $\bar{x}_*(t)$ and $\bar{L}(t)$ are, respectively, the mean of $x_*(t)$ and $\hat{L}(t)$. It can be seem that a realisation of $\hat{L}(t)$ is comparable to its mean.  

We repeat the numerical experiment for $d = 40$ and $a = a_1$. Fig.~\ref{fig:LLR_approx_2} shows a realisation of $L(t)$, a realisation of $\hat{L}(t)$, mean of $|L(t) - \hat{L}(t)|$ over 100 realisations, as well as the mean of $\hat{L}(t)$. We can see the approximation holds up till time $t = 40$, which is $\min\{d,d_1\}$. The purpose of this example is to illustrate why we need to include the condition $t \leq \min\{d,d_1\}$. This is because $L(t)$ and $\hat{L}(t)$ behave differently for $t > \min\{d,d_1\}$ if $d < d_1$. For $L(t)$, it falls after $t = 40$ because from this time onwards, the input signal $s(t)$ being used is small; this leads to a small number of activations and consequently a negative RHS for Eq.~(\ref{eq:L}). However, for $\hat{L}(t)$, the RHS of Eq.~\eqref{eq:Lfinal} is zero because a small $s(t)$ makes $[\phi(s(t))]_+$ zero. 

We have so far used $a = a_1$ and two different durations $d$. We now illustrate that the approximation holds for a different amplitude $a$. For the next numerical experiments, we keep $d = 40$ and use $a =  37.5$. The results are shown in Fig.~\ref{fig:LLR_approx_3}. We can see the approximation holds up till time $\min\{d,d_1\}$. 

These examples demonstrate that, for persistent signals, the approximation $\hat{L}(t) \approx L(t)$ holds for different values of input duration $d$ and amplitude $a$. 

We also want to point out that the behaviour of $\hat{L}(t)$ for transient and persistent signals is consistent with that of the ideal C1-FFL discussed in Section \ref{sec:bg:c1ffl}. We have already pointed out that this is true for transient signals. For a persistent signals, $\hat{L}(t)$ is zero initially and then followed by a non-zero output, i.e. there is a delay before $\hat{L}(t)$ becomes positive and this also holds for the ideal C1-FFL, see the bottom plot in Fig.~\ref{fig:bg:persistent}. We will now map the intermediate approximation Eq.~(\ref{eq:Lfinal}) to the reaction-rate equations of a C1-FFL. 


{\color{black} 
\begin{remark}

We want to remark that in the above formulation and numerical examples, the input signal $s(t)$ is allowed to differ from the two reference signals $c_0(t)$ and $c_1(t)$. Since the decision of the detection problem is based on the log-likelihood ratio in Eq.~\ref{eq:LLR_t}, we can interpret the detection problem as using the history ${\cal X}_*(t)$ (which is generated by $s(t)$) to decide which of the two signals $c_0(t)$ and $c_1(t)$ is more likely to have produced the observed history. Furthermore, consider the case that $s(t)$ is parameterised by positive parameters $a$ and $d$ as in \eqref{eq:def_pulse_st}, then it can be shown that a small change in $a$ or $d$ will be produce a small change in the mean of $L(t)$ and $\hat{L}(t)$. 
\end{remark}  
}


\subsection{Using C1-FFL to approximately compute $\hat{L}(t)$} 
The aim of this section is to show that the C1-FFL can be used to approximately compute the intermediate approximation $\hat{L}(t)$ in Eq.~(\ref{eq:Lfinal}). Recall that the C1-FFL in Fig.~\ref{fig:c1ffl} transforms the signal $x_*(t)$ into the output signal $z(t)$ using the the following components: Nodes Y and Z, and the AND logic. We will model these components using the following chemical reaction system: 
\begin{subequations}
\label{eq:ffl_all}
\begin{eqnarray}
\frac{dy(t)}{dt} &=&  \underbrace{\frac{h_y x_*(t)^{n_y}}{K_{y}^{n_y} + x_*(t)^{n_y}}}_{H_y(x*(t))} - d_y y(t) \label{eq:ffl2} \\
\frac{dz(t)}{dt} &=&  x_*(t)  \times \underbrace{\frac{h_z y(t)^{n_z}}{K_{z}^{n_z} + y(t)^{n_z}}}_{H_z(y(t))}   \label{eq:ffl3} 
\end{eqnarray}
\end{subequations}
where $h_y$, $n_y$, $K_y$ etc.~are coefficients of the Hill functions. We assume that the initial conditions are $y(0) = z(0) = 0$. Note that these two equations are comparable to the ideal C1-FFL model in Section \ref{sec:bg:c1ffl}. In particular, if we replace the $\theta$-function in \eqref{eq:c1ffl:ideal1} by a Hill function, then it becomes  \eqref{eq:ffl2}. Also, if we choose $K_{xz} = 0$ and $\alpha_z = 0$, and replace the $\theta$-function in $z(t)$ by a Hill function in \eqref{eq:c1ffl:ideal2}, then it becomes \eqref{eq:ffl3}. 

By comparing the RHSs of the Eq.~(\ref{eq:Lfinal}) and \eqref{eq:ffl3}, we see that the intermediate approximation $\hat{L}(t)$ and the output of the C1-FFL $z(t)$ can be made approximately equal if $k_- \pi(t) [\phi(s(t))]_+ (=\eta(t))$ in \eqref{eq:Lfinal} and $H_z(y(t))$ in \eqref{eq:ffl3} are approximately equal. We argue in Appendix \ref{app:ia2c1ffl} that it is possible to choose the parameters in \eqref{eq:ffl_all} such that $\eta(t) \approx H_z(y(t))$ in the time interval $[0,\min\{d,d_1\})$. The argument consists of two parts, for the two time intervals $[0,d_0)$ and $[d_0,\min\{d,d_1\})$. 

A major argument made in Appendix \ref{app:ia2c1ffl} is to match $\eta(t)$ and $H_z(y(t))$ in the time interval $[d_0,\min\{d,d_1\})$ for persistent signals. We show in Appendix \ref{app:ia2c1ffl} that this matching problem can be reduced to choosing the parameters in \eqref{eq:ffl_all} so that the following two functions in $a$: $k_- [\phi(a)]_+$ and $H_z(\frac{1}{d_y} H_y(\frac{M k_+ a}{k_+ a + k_-}))$ are approximately equal for a large range of $a$ where $a$, as defined in Section \ref{sec:conn:ia}, is the amplitude of the input $s(t)$ when it is ON. We note in Appendix \ref{app:ia2c1ffl} that these two functions in $a$ can fit to each other because of monotonicity and concavity properties.  

\begin{figure*}[t]
    \centering
    \begin{subfigure}[t]{0.45\textwidth}
        \centering
        \includegraphics[scale=0.35]{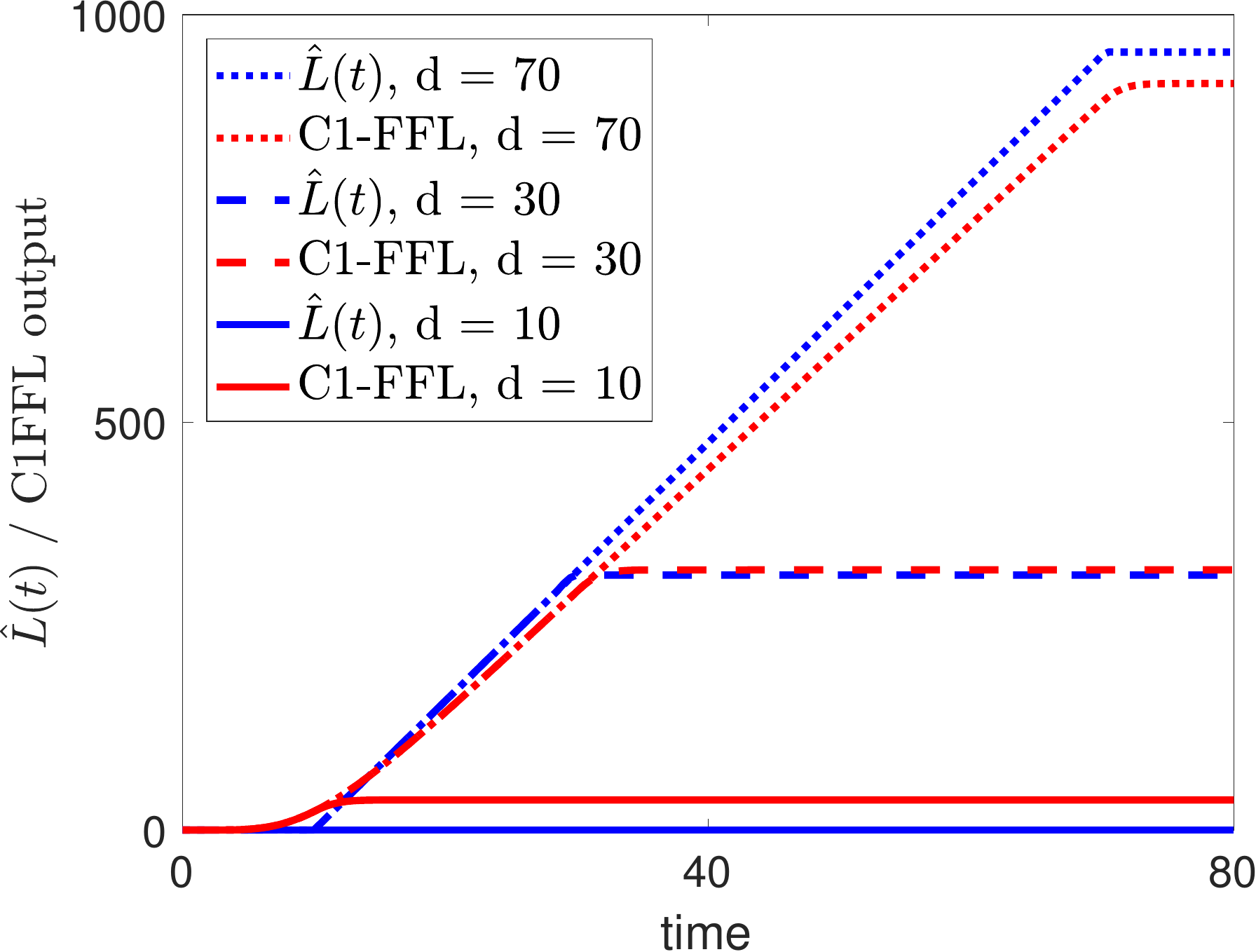}
        \caption{Comparing $\hat{L}(t)$ to C1-FFL output. $a = 5.4.$}
        \label{fig:result1a}
    \end{subfigure}   
    \begin{subfigure}[t]{0.45\textwidth}
        \centering
        \includegraphics[scale=0.35]{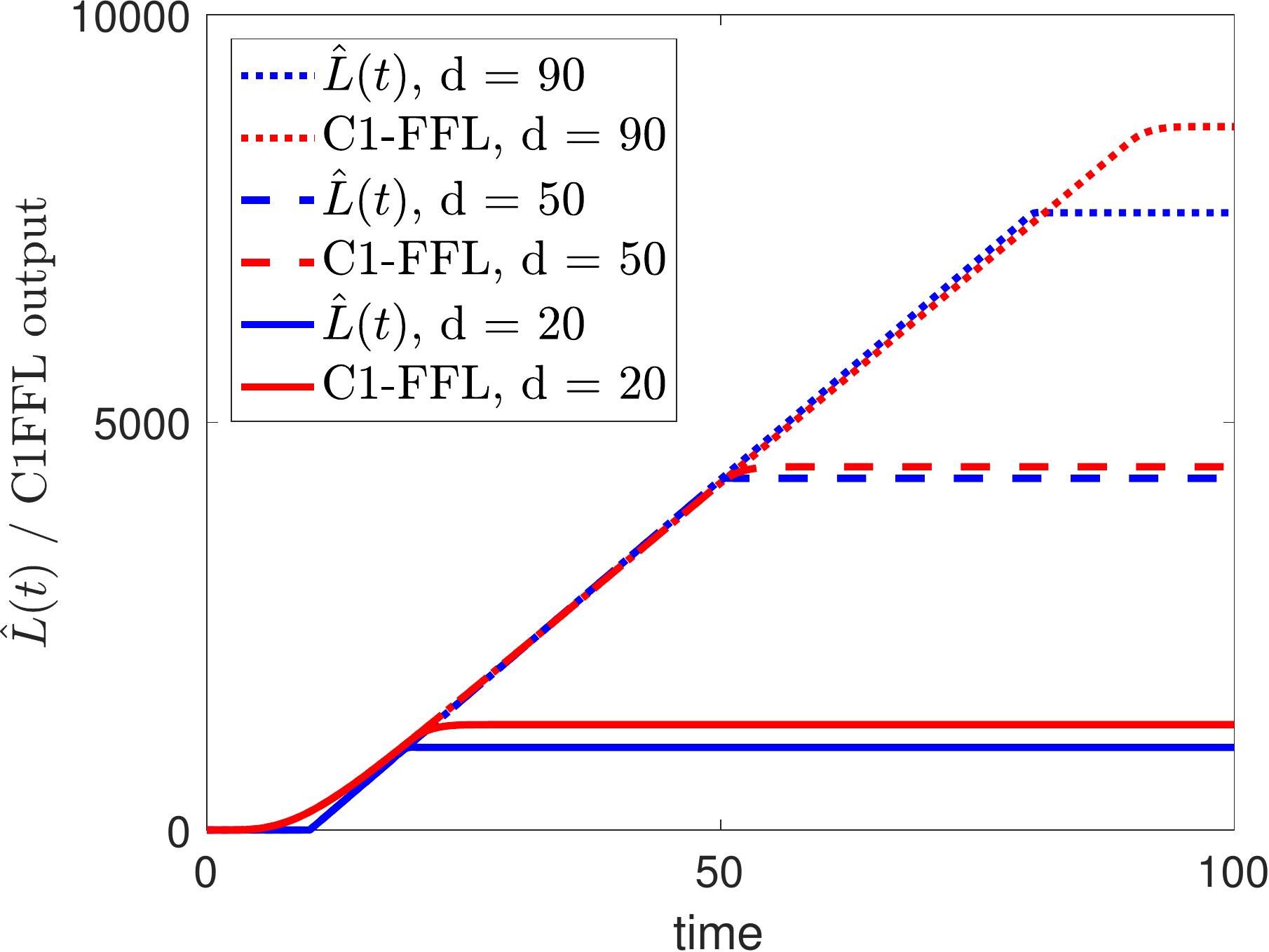}
        \caption{Comparing $\hat{L}(t)$ to C1-FFL output. $a = 40.2.$}
        \label{fig:result1b}
    \end{subfigure}  
    
    \begin{subfigure}[t]{0.45\textwidth}
        \centering
        \includegraphics[scale=0.35]{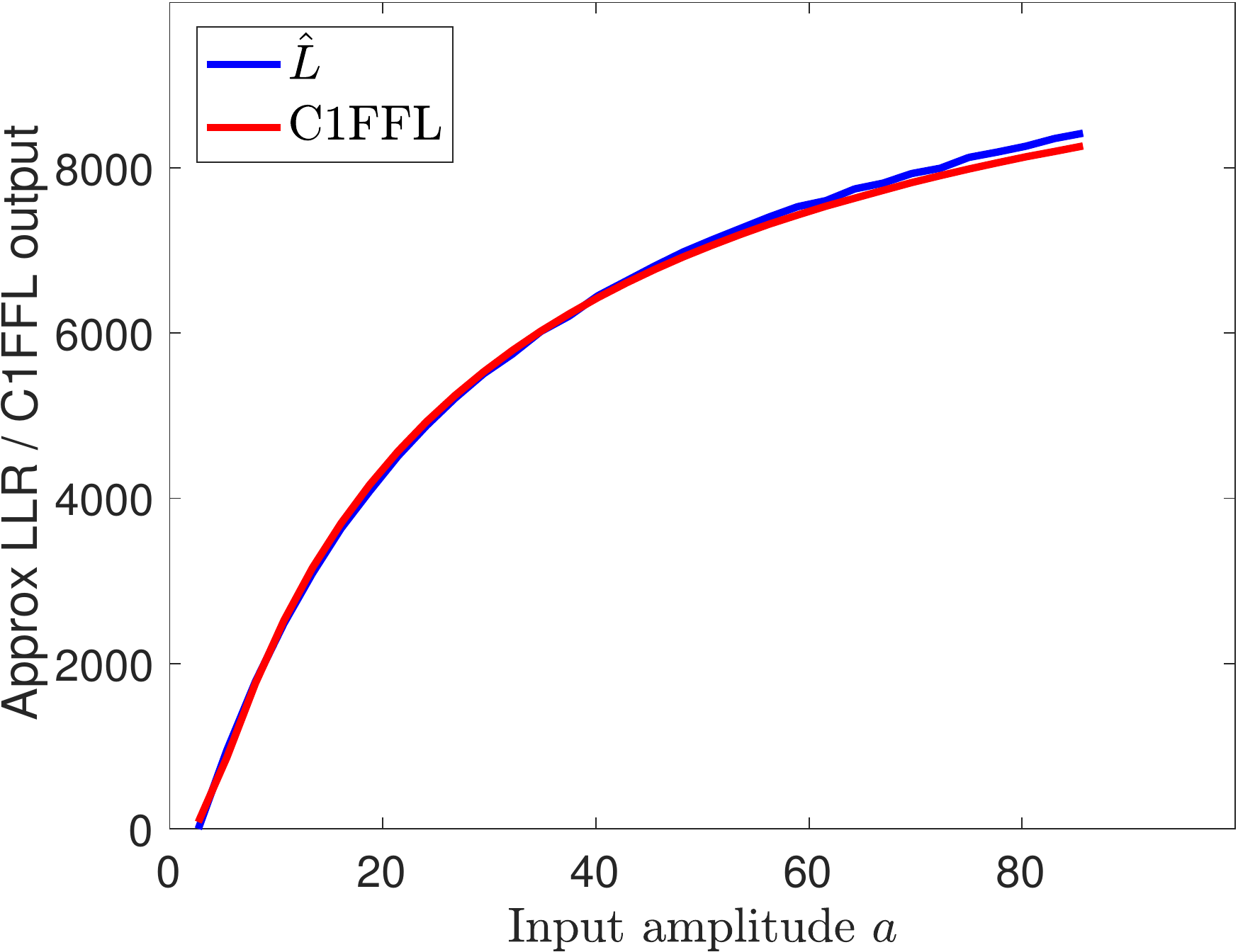}
        \caption{Comparing $\hat{L}(t)$ and C1-FFL output for different pulse input amplitudes.}
        \label{fig:result2}
    \end{subfigure}
        \begin{subfigure}[t]{0.45\textwidth}
        \centering
        \includegraphics[scale=0.35]{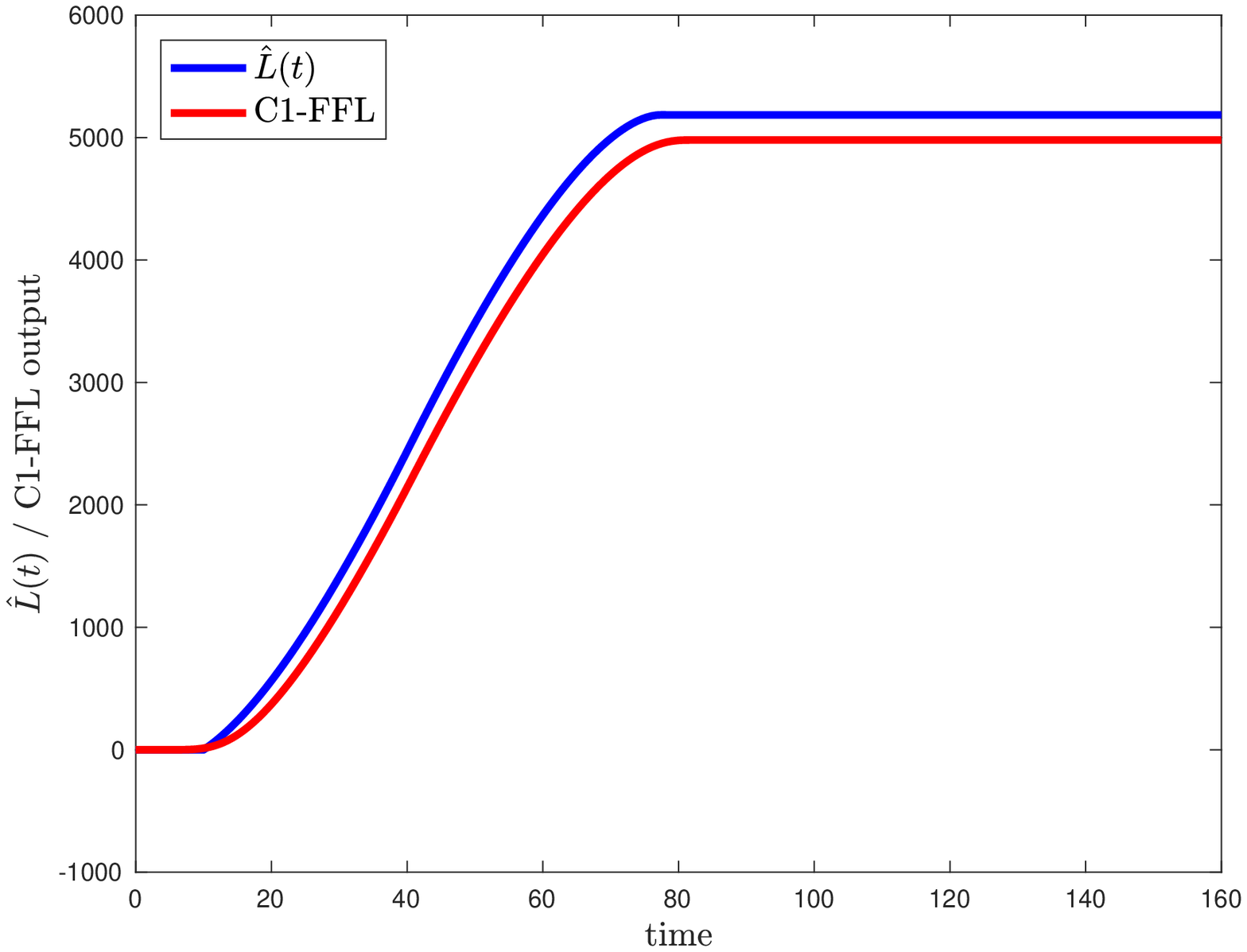}
        \caption{Comparing $\hat{L}(t)$ and C1-FFL output for a triangular pulse.}
        \label{fig:result_tri}
    \end{subfigure}
\caption{Numerical results on C1-FFL. (Best view in color.)}
\end{figure*}

\subsection{Numerical examples} 
We now present numerical examples to show that C1-FFL can be used to compute $\hat{L}(t)$. We use the same $k_+$, $k_-$, $M$, $a_0$ and $a_1$ values as in Section \ref{sec:lhat:eval}. We choose $d_0=10$ and $d_1=80$. We use parameter estimation to determine the parameters in Eq.~(\ref{eq:ffl_all}) so that the C1-FFL output $z(t)$ matches $\hat{L}(t)$ for a range of $a$. {\color{black} The estimated parameters for the C1-FFL are: $h_y = 1.01$, $K_y = 8.04$,  $n_y = 2.26$, $d_y = 0.24$, $h_z = 10.6$, $n_z  = 5.84$ and $K_z = 5.43$.} In this section, we will compare $\hat{L}(t)$ from \eqref{eq:Lfinal} against $z(t)$ from \eqref{eq:ffl_all} assuming the $x_*(t)$ in these two equations is given by $\bar{x}_*(t)$ in \eqref{eq:xstar_bar}. 

Fig.~\ref{fig:result1a} compares $\hat{L}(t)$ and $z(t)$ for input $s(t)$ with $a = 5.4$ and three different durations $d=$ 10, 30 and 70. When $d =10$, the output of the C1-FFL is small. For $d = 30$ and $70$, the C1-FFL output matches well with $\hat{L}(t)$. To show that the match is also good for a different value of the input amplitude $a$, we show the results for $a = 40.2$ and $d=$ 20, 40 and 90 in Fig.~\ref{fig:result1b}. For the case of $d = 90$, we see that the match is good till $[0,d_1)$ because $d > d_1$. 

We have demonstrated that $z(t)$ matches $\hat{L}(t)$ for two different values of $a$. We can show that the match is good for a large range of $a$. We fix the duration $d$ to be $70$ but vary the amplitude $a$ from 2.7 to 85.7. Fig.~\ref{fig:result2} compares $\hat{L}(t)$ and $z(t)$ at $t = 70$. It can be seen that the C1-FFL approximation works for a wide range of $a$. 

The previous examples show that we can match the C1-FFL output $z(t)$ to the intermediate approximation $\hat{L}(t)$ for pulse input $s(t)$ of different durations and amplitudes. We can also show that the match extends to slowly-varying inputs. In this example, we assume $s(t)$ is a triangular pulse with $s(0) = 0$ and rises linearly to $s(40)=42.8$ and then decreases linearly to $s(80)=0$. Fig.~\ref{fig:result_tri} shows the time responses $z(t)$ and $\hat{L}(t)$, and they are comparable. 

\begin{remark}
We finish this section by making {\color{black} a number of} remarks. 
\begin{itemize}
\item Note that we have not included the degradation of Z in \eqref{eq:ffl3} so that we can match it to $\hat{L}(t)$, which does not decay. It can be shown that if we add a degradation term $-\alpha z(t)$ to the RHS of \eqref{eq:ffl3} and $-\alpha \hat{L}(t)$ to the RHS of Eq.~(\ref{eq:Lfinal}), the resulting $z(t)$ will still be matched to $\hat{L}(t)$. 
\item Eq.~(\ref{eq:ffl3}) is not the most general form of C1-FFL. In the general form of Eq.~(\ref{eq:ffl3}), which is presented in \cite{Mangan:2003ja}, the factor $x(t)$ is replaced by a Hill function of $x(t)$. We conjecture that it is possible to generalise the methodology in this paper to obtain the general case and we leave it as future work.  
\item {\color{black} The intermediate approximation $\hat{L}(t)$ is derived under the assumption that $s(t)$ is a rectangular pulse. Future work is needed to better understand the behaviour the intermediate approximation $\hat{L}(t)$ when this assumption does not hold.}
\item  {\color{black}  Although we have shown that the approximate positive log-likelihood ratio in Eq.~(\ref{eq:Lfinal}) can be computed by a C1-FFL, it is certainly not true that any C1-FFL can be used to realise Eq.~(\ref{eq:Lfinal}).  This can be seen from the fact that the C1-FFL model in \eqref{eq:ffl_all} has 7 parameters while the log-likelihood ratio calculation in Eq.~(\ref{eq:Lfinal}) has only 4 parameters. A research question is whether any C1-FFL that can detect persistent signals has a corresponding log-likelihood ratio detector Eq.~(\ref{eq:Lfinal}). We can answer this question by first characterising the C1-FFL that can detect persistent signals and check whether such a correspondence exists. This is an open research problem to be addressed. } 
\item  {\color{black}  We have so far assumed that $c_0(t)$ and $c_1(t)$ are strictly positive for all $t$ by assuming that $a_0 > 0$. If $a_0 = 0$, then the log-likelihood ratio is no longer well-defined because both \eqref{eq:LLR_t} and \eqref{eq:Lfinal} diverge. However, we can compute a shifted and scaled version of the log-likelihood ratio whose intermediate approximation for persistent signals is:
\begin{eqnarray}
\frac{d\hat{L}(t)}{dt} &=& \; x_*(t) \times  \pi(t) \nonumber
\end{eqnarray}
It is still possible to use this intermediate approximation to detect persistent signals. This intermediate approximation can also be approximated by a C1-FFL. Details are omitted and will be studied in future work. 
 } 
\end{itemize} 
\end{remark}

\section{Conclusions and Discussion} 
In this paper, we study the persistence detection property of C1-FFL from an information processing point of view. We formulate a detection problem on a chemical-reaction cycle to understand how an input signal of a long duration can be distinguished from one of short duration. We solve this detection problem and derive an ODE which describes the time evolution of the log-likelihood ratio. An issue with this ODE is that it is difficult to realise it using chemical reactions. We then use time-scale separation to derive an ODE which can approximately compute the log-likelihood ratio when the input signal is persistent. We further show that this approximate ODE can be realised by a C1-FFL. It also provides an interpretation of the persistence detection property of C1-FFL as an approximate computation of log-likelihood ratio. 

The concept of log-likelihood ratio (or a similar quantity) has been used to understand how cells make decision in \cite{Kobayashi:2011dh,Siggia:2013dd}. The paper \cite{Kobayashi:2011dh} considers the problem of distinguishing between two environment states, which are the presence and absence of stimulus. It derives an ODE of the log-odds ratio and uses the ODE to deduce a biochemical network implementation in the form of a a phosphorylation-dephosphorylation cycle. In this cycle, the fraction of phosphorylated substrate is the posteriori probability of the presence of stimuli. The paper \cite{Siggia:2013dd} considers the problem of distinguishing between two different levels of concentration using likelihood ratio. It also presents a molecular implementation that computes the likelihood ratio. This paper differs from \cite{Kobayashi:2011dh,Siggia:2013dd} in one major way. We make a crucial approximation by considering only positive log-likelihood ratio and ignoring negative log-likelihood ratio. We are then able to connect the computation of positive log-likelihood ratio with the computation carried out by a C1-FFL. This work therefore provides a connection between detection theory and C1-FFL using the positive log-likelihood ratio as the connecting point. 

The computation of positive log-likelihood ratio by C1-FFL, which is the key finding of this paper, is an example of using biochemical networks to perform analog computation. There are a few other examples. The incoherent type-1 feedforward loop, which is another network motif, is found to be able to compute fold change \cite{Goentoro:2009gsa}. Allosteric protein is found to be able to compute logarithm approximately \cite{Olsman:2016cr}. In addition, there is also work on using synthetic biochemical circuits to do analog computation \cite{Daniel:2013ke,Chou:2017bx}. 

{\color{black} In this paper, we use a methodology which is based on three key ingredients --- statistical decision theory, time-scale separation and analog molecular computation --- to derive a molecular network that can be used to discriminate persistent signals from transient ones. A possible application of the methodology of this paper in molecular biology is to derive the molecular networks that can decode temporal signals. According to the review paper on temporal signals in cell signalling \cite{Purvis:2013dd}, only some of the molecular networks for decoding temporal signals have been identified. In fact, the authors of \cite{Purvis:2013dd} went further to state that ``Identifying the mechanisms that decode dynamics remains one of the most challenging goals for the field." In \cite{Chou:arxiv_cm}, we used a methodology --- which is similar to the one used in this paper and is based on the same three key ingredients --- to derive a molecular network to decode concentration modulated signals. The derived molecular network was found to be consistent with the Saccharomyces cerevisiae DCS2 promotor data in \cite{Hansen:2013fs}, which were obtained from exciting the promotor by using various transcription factor dynamics, e.g. concentration modulation, duration modulation and others. Another possible application of the methodology of this paper is in synthetic biology. For example, in \cite{Chou:2018fv} we used a methodology --- which is similar to the one used in this paper and in \cite{Chou:arxiv_cm} --- to derive a {\sl de novo} molecular network for decoding concentration modulated signals. We remark that the molecular networks in \cite{Chou:arxiv_cm} and \cite{Chou:2018fv} can be interpreted as an approximate log-likelihood detector of concentration modulated signals. 
}

A recent report \cite{Gerardin:2016fd} considers the problem of determining the biochemical circuits that can be used to distinguish between a persistent and a transient signal. By searching over all biochemical circuits with a limited complexity, the authors find that there are five different circuits that can be used. One of these is C1-FFL. An open question is whether one can use the framework in this paper to deduce all circuits that can detect persistent signals. If this is possible, then it presents an alternative method to find the biochemical circuits that can realise a function. \\

{\color{black}
\subsubsection*{Data availability}
The source code for producing the results for this paper is available at Github, which is an open online code repository. The source code is at \href{https://github.com/ctchou-unsw/c1ffl-journal}{https://github.com/ctchou-unsw/c1ffl-journal} \cite{Chou:c1ffl_code} 
}
\subsubsection*{Acknowledgements} 
The author wishes to thank Dr.~Guy-Bart Stan, Imperial College, for the suggestion to consider possible connections with motifs.


\begin{thebibliography}{10}

\bibitem{Milo:2002cg}
Milo R, Shen-Orr S, Itzkovitz S, Kashtan N, Chklovskii D, Alon U.
\newblock {Network motifs: simple building blocks of complex networks.}
\newblock Science. 2002 Oct;298(5594):824--827.

\bibitem{ShenOrr:2002jo}
Shen-Orr SS, Milo R, Mangan S, Alon U.
\newblock {Network motifs in the transcriptional regulation network of
  Escherichia coli}.
\newblock Nature genetics. 2002 Apr;31(1):64--68.

\bibitem{Alon:2007uu}
Alon U.
\newblock {Network motifs: theory and experimental approaches}.
\newblock Nature Reviews Genetics. 2007;.

\bibitem{Mangan:2003ja}
Mangan S, Alon U.
\newblock {Structure and function of the feed-forward loop network motif.}
\newblock Proceedings of the National Academy of Sciences of the United States
  of America. 2003 Oct;100(21):11980--11985.

\bibitem{Mangan:2003ia}
Mangan S, Zaslaver A, Alon U.
\newblock {The Coherent Feedforward Loop Serves as a Sign-sensitive Delay
  Element in Transcription Networks}.
\newblock Journal of molecular biology. 2003 Nov;334(2):197--204.

\bibitem{Kay_v2}
Kay SM.
\newblock Fundamentals of Statistical Signal Processing, Volume II: Detection
  Theory.
\newblock Prentice Hall; 1998.

\bibitem{Alon}
Alon U.
\newblock An Introduction to Systems Biology: Design Principles of Biological
  Circuits.
\newblock Chapman \& Hall; 2006.

\bibitem{Gardiner}
Gardiner C.
\newblock Stochastic methods.
\newblock Springer; 2010.

\bibitem{Siggia:2013dd}
Siggia ED, Vergassola M.
\newblock {Decisions on the fly in cellular sensory systems.}
\newblock Proceedings of the National Academy of Sciences. 2013
  Sep;110(39):E3704--12.

\bibitem{Kobayashi:2011dh}
Kobayashi TJ, Kamimura A.
\newblock {Dynamics of intracellular information decoding}.
\newblock Physical Biology. 2011 Aug;8(5):055007.

\bibitem{Chou:gc}
Chou CT.
\newblock {Maximum a-posteriori decoding for diffusion-based molecular
  communication using analog filters}.
\newblock Nanotechnology, IEEE Transactions on. 2015;14(6):1054--1067.

\bibitem{Gillespie:1977ww}
Gillespie D.
\newblock {Exact stochastic simulation of coupled chemical reactions}.
\newblock The journal of physical chemistry. 1977;.

\bibitem{Oishi:2011ig}
Oishi K, Klavins E.
\newblock {Biomolecular implementation of linear I/O systems}.
\newblock Systems Biology, IET. 2011 Jul;5(4):252--260.

\bibitem{Goentoro:2009gsa}
Goentoro L, Shoval O, Kirschner MW, Alon U.
\newblock {The Incoherent Feedforward Loop Can Provide Fold-Change Detection in
  Gene Regulation}.
\newblock Molecular cell. 2009 Dec;36(5):894--899.

\bibitem{Olsman:2016cr}
Olsman N, Goentoro L.
\newblock {Allosteric proteins as logarithmic sensors.}
\newblock Proceedings of the National Academy of Sciences. 2016 Jul;p.
  201601791.

\bibitem{Daniel:2013ke}
Daniel R, Rubens JR, Sarpeshkar R, Lu TK.
\newblock {Synthetic analog computation in living cells}.
\newblock Nature. 2013 May;497(7451):619--623.

\bibitem{Chou:2017bx}
Chou CT.
\newblock {Chemical reaction networks for computing logarithm}.
\newblock Synthetic Biology. 2017 Apr;2(1):1--13.

\bibitem{Purvis:2013dd}
Purvis JE, Lahav G.
\newblock {Encoding and Decoding Cellular Information through Signaling
  Dynamics}.
\newblock Cell. 2013 Feb;152(5):945--956.

\bibitem{Chou:arxiv_cm}
Chou CT.
\newblock Designing molecular circuits for approximate maximum a posteriori
  demodulation of concentration modulated signals.
\newblock arXiv; 2018. arXiv:1808.01543.

\bibitem{Hansen:2013fs}
Hansen AS, O'Shea EK.
\newblock {Promoter decoding of transcription factor dynamics involves a
  trade-off between noise and control of gene expression}.
\newblock Molecular systems biology. 2013 Nov;9:1--14.

\bibitem{Chou:2018fv}
Chou CT.
\newblock {Molecular circuit for approximate maximum a posteriori demodulation
  of concentration modulated signals}.
\newblock In: the 5th ACM International Conference on Nanoscale Computing and
  Communication. New York, New York, USA: ACM; 2018. .

\bibitem{Gerardin:2016fd}
Gerardin J, Lim WA.
\newblock {The design principles of biochemical timers: circuits that
  discriminate between transient and sustained stimulation}.
\newblock bioRxiv. 2016 May;p. 1--51.

\bibitem{Chou:c1ffl_code}
\url{https://github.com/ctchou-unsw/c1ffl-journal};.

\bibitem{Grimmett}
Grimmett GR, Stirzaker DR.
\newblock Probability and Random Processes.
\newblock Oxford University Press; 1997.

\end{thebibliography}

\appendix
\section{Proof and derivation}

\subsection{Proof of (\ref{eq:L})} 
\label{app:sol:dp}
Recalling that ${\cal X}_*(t)$ is the history of $x_*(t)$ in the time interval $[0,t]$. In order to derive (\ref{eq:L}), we consider the history ${\cal X}_*(t+\Delta t)$ as a concatenation of ${\cal X}_*(t)$ and $x_*(t)$ in the time interval $(t,t+\Delta t]$. We assume that $\Delta t$ is chosen small enough so that no more than one activation or deactivation reaction can take place in $(t, t+\Delta t]$. Given this assumption and right continuity of continuous-time Markov Chains, we can use $x_*(t+\Delta t)$ to denote the history of $x_*(t)$ in $(t,t+\Delta t]$. 

Consider the likelihood of observing ${\cal X}_*(t+\Delta t)$ given hypothesis ${\cal H}_i$: 
\begin{eqnarray}
&     & {\rm P}[{\cal X}_*(t+\Delta t) | {\cal H}_i]  \label{eqn:star:like:1} \\
& = & {\rm P}[{\cal X}_*(t) \; \mbox{\textsc{and}} \; x_*(t+\Delta t) | {\cal H}_i] \label{eqn:star:like:2} \\
& = & {\rm P}[{\cal X}_*(t) | {\cal H}_i]  \; {\rm P}[ x_*(t+\Delta t) | {\cal H}_i, {\cal X}_*(t)] \label{eqn:star:like:3} \\
& = & {\rm P}[{\cal X}_*(t) | {\cal H}_i]  \; {\rm P}[ x_*(t+\Delta t) | {\cal H}_i, x_*(t)]  \label{eqn:star:like:4}
\end{eqnarray}
where we have expanded ${\cal X}_*(t+\Delta t)$ in Eq.\eqref{eqn:star:like:1} using concatenation and used Markov property to go from Eq.\eqref{eqn:star:like:3} to Eq.\eqref{eqn:star:like:4}. 

By using \eqref{eqn:star:like:4} in the definition of log-likelihood ratio, we can show that: 
\begin{align}
L(t+\Delta t) = L(t) + \log \left( \frac{{\rm P}[x_*(t+\Delta t) | {\cal H}_1, x_*(t)]}{{\rm P}[x_*(t+\Delta t) | {\cal H}_0, x_*(t)]} \right) \label{eq:app:L}
\end{align}

The value of the expression ${\rm P}[x_*(t+\Delta t) | {\cal H}_i, x_*(t)]$ depends on whether $x_*(t+\Delta t)$ is one greater than, one less than or equal to $x_*(t)$. These cases correspond, respectively, to the event that an X molecule been activated, an X$_*$ molecule been deactivated and no change in the state of the molecules in the time interval $(t,t+\Delta t]$. Under the hypothesis ${\cal H}_i$, which means the input signal is assumed to be $c_i(t)$, the activation and deactivation rates are, respectively, $k_+ (M - x_*(t))  c_i(t)$  and $k_- x_*(t)$ when the number of X$_*$ molecules is $x_*(t)$. We can therefore write the expression ${\rm P}[x_*(t+\Delta t) | {\cal H}_i, x_*(t)]$ as: 
\begin{align}
&  {\rm P}[x_*(t+\Delta t) | {\cal H}_i, x_*(t)] =\nonumber \\ 
& \delta_{x_*(t+\Delta t),x_*(t) + 1} k_+ (M - x_*(t))  \;  c_i(t) \; \Delta t +  \nonumber \\ 
& \delta_{x_*(t+\Delta t),x_*(t) - 1} k_- x_*(t) \; \Delta t \; + \nonumber  \\
& \delta_{x_*(t+\Delta t),x_*(t)} (1 - k_+ (M - x_*(t)) c_i(t) \; \Delta t - k_- x_*(t)  \; \Delta t) \label{eq:app:predictb}
\end{align}
where $\delta_{a,b}$ is the Kronecker delta which is 1 when $a = b$. 

Note that ${\rm P}[x_*(t+\Delta t) | {\cal H}_i, x_*(t)]$ in \eqref{eq:app:predictb} is a sum of three terms with multipliers $\delta_{x_*(t+\Delta t),x_*(t) + 1}$, $\delta_{x_*(t+\Delta t),x_*(t) - 1}$ and $\delta_{x_*(t+\Delta t),x_*(t)}$. Since these multipliers are mutually exclusive, we have:
\begin{align}
 & \log \left( \frac{{\rm P}[x_*(t+\Delta t) | {\cal H}_1, x_*(t)]}{{\rm P}[x_*(t+\Delta t) | {\cal H}_0, x_*(t)]} \right)   \nonumber  \\
 = & 
 \delta_{x_*(t+\Delta t),x_*(t) + 1} \log \left( \frac{k_+ (M - x_*(t))  \;  c_1(t) \; \Delta t  }{k_+ (M - x_*(t))   \;  c_0(t) \; \Delta t }  \right) + \nonumber \\
 & \delta_{x_*(t+\Delta t),x_*(t) - 1} \log \left( \frac{k_- x_*(t) \; \Delta t}{k_- x_*(t) \; \Delta t}   \right) \nonumber + \\
 & \delta_{x_*(t+\Delta t),x_*(t)} \log \left( \frac{ 1 - k_+ (M - x_*(t)) c_1(t) \; \Delta t - k_- x_*(t)  \; \Delta t}{ 1 - k_+ (M - x_*(t)) c_0(t) \; \Delta t - k_- x_*(t)  \; \Delta t }   \right) \nonumber \\
 \approx & 
\delta_{x_*(t+\Delta t),x_*(t) + 1} \log \left( \frac{c_1(t) }{c_0(t) }  \right) - \nonumber \\
& \delta_{x_*(t+\Delta t),x_*(t)} k_+ (M - x_*(t))  
\left( c_1(t) - c_0(t)  \right) \; \Delta t \label{eqn:app:likelihood} 
\end{align}
where we have used the approximation $\log(1 + f \; \Delta t) \approx f \; \Delta t$ to obtain \eqref{eqn:app:likelihood}. {\color{black} Note also that the above  derivation requires that both $c_0(t)$ and $c_1(t)$ must be strictly positive for all $t$.}

By substituting Eq.~\eqref{eqn:app:likelihood} into Eq.~(\ref{eq:app:L}), we have after some manipulations and after taking the limit $\Delta t \rightarrow 0$:
\begin{align}
\frac{dL(t)}{dt} 
= & \lim_{\Delta t \rightarrow 0} \frac{\delta_{x_*(t+\Delta t),x_*(t) + 1} }{\Delta t}
\log \left( \frac{c_1(t)}{c_0(t)}  \right) - \nonumber \\
& \delta_{x_*(t+\Delta t),x_*(t)} k_+ (M - x_*(t))  
\left( c_1(t) - c_0(t)  \right) \label{eqn:app:logmapm1}  
\end{align}
In order to obtain Eq.~(\ref{eq:L}), we use the following reasonings. First, the term $\lim_{\Delta t \rightarrow 0} \frac{\delta_{x_*(t+\Delta t),x_*(t) + 1} }{\Delta t}$ is a Dirac delta at the time instant that an \cee{X} molecule is activated. Second, the term $\delta_{x_*(t+\Delta t),x_*(t)}$ is only zero when the number of X$_*$ molecule changes but the number of such changes is countable. In other words, $\delta_{x_*(t+\Delta t),x_*(t)}=1$ with probability one. This allows us to drop $\delta_{x_*(t+\Delta t),x_*(t)}$. Hence Eq.~(\ref{eq:L}).  

\subsection{Derivation of \eqref{eq:Lfinal}}
\label{app:Lhat}
The aim of this appendix is to derive the intermediate approximation \eqref{eq:Lfinal}. We will split the derivations into two parts, depending on the length of the duration $d$ relative to $d_0$. We first consider the case where the input signal $s(t)$ has a duration longer than or equal to $d_0$, which is also the more important case for the derivation because it deals with persistent signals. 

Our aim is to find an approximation of log-likelihood ratio $L(t)$ given in \eqref{eq:L}. Our strategy is to divide time into intervals such that, in each time interval, each of the time profiles of $c_0(t)$, $c_1(t)$ and $s(t)$ is a constant. 

The first time interval is $[0,d_0)$ where $c_0(t) = c_1(t) = a_1$ and $s(t) = a$. Since $L(0) = 0$ and $\frac{dL(t)}{dt} = 0$ in this time interval, therefore $L(t) = 0$ in this time interval. 

The next time interval to consider  is $[d_0,\min\{d,d_1\})$ where $c_0(t) = a_0$, $c_1(t) = a_1$ and $s(t) = a$. For $t \in [d_0,\min\{d,d_1\})$, the log-likelihood ratio $L(t)$ in \eqref{eq:L} can be written as $L(t) = L_1(t) + L_2(t)$ where
\begin{eqnarray}
L_1(t) & = & \log\left(\frac{a_1}{a_0} \right) \underbrace{\int_{d_0}^t \left[ \frac{dx_*(t)}{dt} \right]_+   dt}_{A(t)}  \label{eq:app_L1} \\
L_2(t) & = & - k_+ (a_1- a_0) \int_{d_0}^t  (M - x_*(t))  dt . \label{eq:app_L2}
\end{eqnarray}  

We first consider finding an approximation of the integral $A(t)$ in \eqref{eq:app_L1} and the aim is to replace the positive derivative of $x_*(t)$ by some other arithmetic operations which can be computed by using chemical reactions. The integral $A(t)$ can be interpreted as the number of times that ${\cee X}$ is activated in the time interval $[d_0,t)$. For an ${\cee X}$ molecule, the time between two consecutive activations is a random variable with mean $m$ and variance $\sigma^2$ where: 
\begin{align}
m =& \frac{1}{k_+ a} + \frac{1}{k_-} \label{eq:app:mean_activation} \\
\sigma^2 = & \frac{1}{(k_+ a)^2} + \frac{1}{k_-^2} 
\end{align}
This is because we can model the activation and deactivation of ${\cee X}$ by a 2-state continuous-time Markov chain with transition rates $k_+ a$ and $k_-$. 

We will now make a time-scale separation assumption by assuming that the duration $(t-d_0)$ is much bigger than $m$, i.e. $t - d_0 \gg \frac{1}{k_+ a} + \frac{1}{k_-}$. This assumption can be met by having a sufficiently long duration $d$ and large amplitude $a$. If this time-scale separation assumption holds, then there are many activations in the time interval $[d_0,t)$. In this case, we can use the renewal theorem \cite{Grimmett} to approximate $A(t)$, we have:
\begin{eqnarray}
{\rm mean}(A(t)) & \approx & M \frac{t-d_0}{m}  \\
{\rm var}(A(t)) & \approx &M \frac{\sigma^2}{m^3} (t-d_0),    
\end{eqnarray}
which implies that  
\begin{align}
\frac{ \sqrt{ {\rm var}(A(t)) } }{ {\rm mean}(A(t)) } \approx& \frac{ \sigma }{  m \sqrt{M} \sqrt{t-d_0} }. 
\end{align}
This means we can approximate $A(t)$ by its mean and the error decreases with the reciprocal of the square root of the duration $t-d_0$. By using this approximation, we have:
\begin{eqnarray}
L_1(t) & \approx & \log\left(\frac{a_1}{a_0} \right) \frac{M}{m} (t-d_0)  \label{eq:app_L1_1} 
\end{eqnarray}  

The time-scale separation assumption also implies that the ensemble average of $x_*(t)$ can be treated as a constant in the time interval $[d_0,t)$; we will denote this average by $x_{*,a}$ where 
\begin{eqnarray}
x_{*,a} & = & \frac{M k_+ a}{k_+ a + k_-} \label{eq:app:ensemble_average} 
\end{eqnarray}
This ensemble average is related to mean inter-activation time $m$ in \eqref{eq:app:mean_activation} by:
\begin{eqnarray}
x_{*,a} & = & \frac{M}{m k_-}
\end{eqnarray}
By using this relationship in \eqref{eq:app_L1_1}, we have:
\begin{eqnarray}
L_1(t) & \approx & k_- \; x_{*,a}  \log\left(\frac{a_1}{a_0} \right) (t-d_0)   \label{eq:app_L1_2} 
\end{eqnarray}  
which means $L_1(t)$ can be computed from the ensemble average $x_{*,a}$. We will return to this expression shortly after studying the approximation of the integral in $L_2(t)$ in \eqref{eq:app_L2}. 

Since the Markov chain describing the reaction cycle of ${\cee X}$ and ${\cee X_*}$ is ergodic, the time average in \eqref{eq:app_L2} can be approximated by its ensemble average. By using the ensemble average $x_{*,a}$ in \eqref{eq:app:ensemble_average}, we can show that: 
\begin{eqnarray}
L_2(t) & \approx & - k_-  \;  x_{*,a} \frac{(a_1- a_0)}{a} (t-d_0)  \label{eq:app_L2_2} 
\end{eqnarray}  
  
Since $L(t) = L_1(t) + L_2(t)$, it follows from \eqref{eq:app_L1_2} and \eqref{eq:app_L2_2} that:
\begin{eqnarray}
L(t) & \approx & k_-  \; x_{*,a}  \left( \log\left(\frac{a_1}{a_0} \right)  - \frac{(a_1- a_0)}{a}    \right)  (t-d_0)     \label{eq:app_L_3} 
\end{eqnarray}  
in the time interval $[d_0,\min\{d,d_1\})$. 

We can re-write the results that we have for the time interval $[0,\min\{d,d_1\})$ in differential form, as follows: 
\begin{eqnarray}
\frac{dL(t)}{dt} & \approx &  k_-  \; x_*(t) \left\{ \log\left(\frac{c_1(t)}{c_0(t)} \right)  -  \frac{(c_1(t) - c_0(t))}{s(t)} \right\} \label{eq:star:L2} 
\end{eqnarray} 

The derivation so far has shown that the ODEs \eqref{eq:star:L2} and \eqref{eq:L} are approximately equal for $t$ in $[0, \min\{d,d_1\} )$. We will consider consider $t \geq \min\{d,d_1\}$. We need to split into two cases: $d \geq d_1$ and $d < d_1$. For the first case, the time interval concerned is $t \geq d$. It can be verified that the RHSs of \eqref{eq:star:L2} and \eqref{eq:L} are both zero for this time interval. Thus, if $d \geq d_1$, then $\hat{L} \approx L(t)$ for all $t$. We will consider the second case, where $d < d_1$, in the next paragraph. 

If $d < d_1$, then the time interval $[d,d_1)$ is non-empty. In this interval, we have $s(t) = a_0$, $c_0(t) = a_0$ and $c_1(t) = a_1$, which means the term in curly brackets in \eqref{eq:star:L2} is equal to $\log\left( \frac{a_1}{a_0} \right) - \frac{a_1 - a_0}{a_0}$. Since $a_1 > a_0$, this term is negative. As a result, this may result in a negative $L(t)$. Although we learn from the research on synthetic analog computation using chemical reactions \cite{Oishi:2011ig} that it is possible to handle negative numbers using chemical reactions, the research also tells us that this is inherently a difficult process and the complexity is high. Therefore, we will use an approximation that does not result in a negative log-likelihood ratio. By adding the $[  \; ]_+$ operator, where $[w]_+ = \max(w,0)$, to the term in curly brackets in \eqref{eq:star:L2}, we arrive at:
\begin{eqnarray}
\frac{d\hat{L}(t)}{dt} 
                       &\approx&  k_-  \; x_*(t) \left\{ \left[ \log\left(\frac{c_1(t)}{c_0(t)} \right)  -  \frac{(c_1(t) - c_0(t))}{s(t)} \right]_+ \right\} \label{eq:star:L3} 
\end{eqnarray} 
The addition of the $[  \; ]_+$ operator does not affect what happens in the time interval $[0,\min\{d,d_1\})$. However, it means that the RHS of \eqref{eq:star:L3} does not equal to the RHS of \eqref{eq:L} in the time interval $[d,d_1)$; in fact, this is the only time interval that the RHSs of \eqref{eq:star:L3} and \eqref{eq:L} are not approximately equal. This also means that, for input signals whose duration $d$ is less than $d_1$, the approximation $\hat{L}(t) \approx L(t)$ only holds in the time interval $[0,\min\{d,d_1\})$. 

Our next step is to show that \eqref{eq:star:L3} can be written as \eqref{eq:Lfinal}. By using the form of $c_0(t)$ and $c_1(t)$, we can show that 
\begin{eqnarray}
\log\left(\frac{c_1(t)}{c_0(t)} \right) & = & \log\left( \frac{a_1}{a_0} \right) \; \pi(t) \label{eq:star:w1}  \\ 
c_1(t) - c_0(t) & = & (a_1 - a_0) \; \pi(t) \label{eq:star:w2}  
\end{eqnarray} 
By substituting \eqref{eq:star:w1} and \eqref{eq:star:w2} into \eqref{eq:star:L3}, we arrive at 
\begin{align}
\frac{d\hat{L}(t)}{dt} =& \; x_*(t) \times \left\{ k_- \; \pi(t) \;  \left[\log \left(\frac{a_1}{a_0} \right)  -  \frac{ a_1 - a_0 }{s(t)} \right]_+ \right\}  \label{eq:star:Lfinal}  
\end{align}
which is the same as \eqref{eq:Lfinal}. 

We conclude the derivation of $\hat{L}(t)$ by showing that $\hat{L}(t) = 0$ for all $t$ for input signals $s(t)$ whose duration $d$ is less than $d_0$. This can be done by showing the RHS of \eqref{eq:Lfinal} is zero for all $t$. Since $\pi(t)$ is only non-zero in the time interval $[d_0,d_1)$, we only have to consider this time interval. In this time interval, we can show that $\left[\log \left(\frac{a_1}{a_0} \right)  -  \frac{ a_1 - a_0 }{s(t)} \right]_+$ is zero because $s(t) = a_0$. 


\subsection{Matching \eqref{eq:Lfinal} to \eqref{eq:ffl_all}}
\label{app:ia2c1ffl}
The aim of this appendix is to explain why it is possible to use the C1-FFL system in \eqref{eq:ffl_all} to realise the intermediate approximation in \eqref{eq:Lfinal}. By comparing the RHSs of the \eqref{eq:Lfinal} and \eqref{eq:ffl3}, our aim is to show that, by using appropriate choice of parameters in \eqref{eq:ffl_all}, $k_- \pi(t) [\phi(s(t))]_+ (=\eta(t))$ and $H_z(y(t))$ can be made to be approximately equal in the time interval $[0,\min\{d,d_1\})$. We will consider the time intervals $[0,d_0)$ and $[d_0,\min\{d,d_1\})$ separately. 

We first consider the time interval $[d_0,\min\{d,d_1\})$. It is sufficient to consider only persistent input signals. Within this time interval, the persistent input $s(t)$ has an amplitude of $a$. Since we assume that the input $s(t)$ is long compared to the time-scale of the activation and deactivation reactions, therefore the mean of $x_*(t)$ is a plateau (see the bottom plot of Fig.~\ref{fig:xstar}) whose height is $\frac{M k_+ a}{k_+ a + k_-}$. Consequently, the time profiles of both $\eta(t)$ and $y(t)$ also contain a period of time that they plateau. The plateau in $\eta(t)$ contributes to the ramp-like increase in $\hat{L}(t)$ in Fig.~\ref{fig:LLR_approx}. 

This means that, if want to match \eqref{eq:Lfinal} and \eqref{eq:ffl_all} in the time interval $[d_0,\min\{d,d_1\})$, we need to match the values of $\eta(t)$ and $H_z(y(t))$ at their plateau. Since the amplitude of the input $s(t)$ when it is ON is $a$, the heights of the plateau of $\eta(t)$ and $H_z(y(t))$ are, respectively, $k_- [\phi(a)]_+ (=f_1(a))$ and $H_z(\frac{1}{d_y} H_y(\frac{M k_+ a}{k_+ a + k_-}))(=f_2(a))$, and we want $f_1(a) \approx f_2(a)$ for as large a range of $a$ as possible. Note that for all $a$ such that $f_1(a) > 0$, both functions $f_1(a)$ and $f_2(a)$ are strictly increasing and both $f_1(\infty)$ and $f_2(\infty)$ are constants. Therefore, we can choose the Hill function coefficients to fit $f_2(a)$ to $f_1(a)$. This argument takes care of the case when $s(t)$ is a persistent input which requires us to implement the function $\phi(.)$ in \eqref{eq:Lfinal} using the Hill functions in \eqref{eq:ffl_all}. {\color{black} We remark that we need to include the requirement $f_1(a) > 0$ in the above argument because $f_1(a)$ is not strictly increasing when $f_1(a) = 0$. This can be seen from the fact that there is a range of $a$ such that $\phi(a) < 0$, which means that there is a range of $a$ such that $f_1(a) = 0$, which in turn means that $f_1(a)$ is not monotonically increasing in this range. } 

We now consider the time interval $[0,d_0)$. In this time interval, $\hat{L}(t) = 0$ due to $\pi(t)$. This is a feature shared by the ideal C1-FFL model in \cite{Alon}. The book~\cite{Alon} shows that this can be realised by choosing a big enough $K_z$ in Eq.~(\ref{eq:ffl3}) so that the production rate of $z(t)$ is small initially.

\end{document}